\documentclass[aps,prx,twocolumn,amssymb,amsfonts,superscriptaddress]{revtex4-2}  

\usepackage{natbib}
\usepackage{graphicx}
\usepackage{amsmath}
\usepackage[inline]{enumitem}
\usepackage{textcomp} 
\usepackage[mathscr]{euscript}
\usepackage{bbm}
\usepackage{multirow}

\usepackage{setspace} 
\usepackage{xcolor}

\usepackage{xr}

\def\in{_\text{in}}
\def\out{_\text{out}}
\def\p{_\text{p}}

\def\rout{\mathbf{r\out}}
\def\rin{\mathbf{r\in}}
\def\rp{\mathbf{r\p}}

\def\tin{\theta\in}

\newcommand*{\wil}[1]{\textcolor{black}{#1}}

\newcommand*{\alex}[1]{\textcolor{black}{#1}}

\newcommand*{\al}[1]{\textcolor{black}{#1}}
\newcommand*{\la}[1]{\textcolor{black}{#1}}
\newcommand*{\wi}[1]{\textcolor{black}{#1}}
\newcommand*{\rev}[1]{\textcolor{black}{#1}}

\newcommand*{\lc}[1]{\textcolor{black}{#1}}
\newcommand*{\revv}[1]{\textcolor{black}{#1}}

\begin{document}

\title{Ultrasound Matrix Imaging—Part II: The distortion matrix for aberration correction over multiple isoplanatic patches.}

\author{William Lambert}
\affiliation{Institut Langevin, ESPCI Paris, CNRS, PSL University, 1 rue Jussieu, 75005 Paris, France}
	\affiliation{Hologic / SuperSonic Imagine, 135 Rue Emilien Gautier, 13290 Aix-en-Provence, France}
\author{Laura A. Cobus}
\affiliation{Institut Langevin, ESPCI Paris, CNRS, PSL University, 1 rue Jussieu, 75005 Paris, France}
\affiliation{Dodd-Walls Centre for Photonic and Quantum Technologies and Department of Physics, University of Auckland, Private Bag 92019, Auckland 1010, New Zealand}
\author{Justine Robin}
\affiliation{Institut Langevin, ESPCI Paris, CNRS, PSL University, 1 rue Jussieu, 75005 Paris, France}
\affiliation{Physics for Medicine Paris, INSERM, CNRS, ESPCI Paris, PSL University, 17 rue Moreau, 75012 Paris, France}
\author{Mathias Fink}
\affiliation{Institut Langevin, ESPCI Paris, CNRS, PSL University, 1 rue Jussieu, 75005 Paris, France}
\author{{Alexandre Aubry}}
\email{alexandre.aubry@espci.fr}
\affiliation{Institut Langevin, ESPCI Paris, CNRS, PSL University, 1 rue Jussieu, 75005 Paris, France}





\begin{abstract}
This is the second article in a series of two which report on a matrix approach for ultrasound imaging in heterogeneous media. This article describes the quantification and correction of aberration, i.e. the distortion of an image caused by spatial variations in the medium speed-of-sound. Adaptive focusing can compensate for  aberration, but is only effective over a restricted area called the isoplanatic patch. Here, we use an experimentally-recorded matrix of reflected acoustic signals to
synthesize a set of virtual transducers. We then examine wave propagation between these virtual transducers and an arbitrary correction plane.  Such wave-fronts consist of two components: (\textit{i}) An ideal geometric wave-front linked to diffraction and the input focusing point, and; (\textit{ii}) Phase distortions induced by the speed-of-sound variations. These distortions are stored in
 a so-called distortion matrix, the singular value decomposition of which gives access to an optimized focusing law at any point. We show that, by decoupling the aberrations undergone by the outgoing and incoming waves and applying an iterative strategy, compensation for even high-order and spatially-distributed aberrations can be achieved. After a numerical validation of the process, ultrasound matrix imaging (UMI) is applied to the \textit{in-vivo} imaging of a gallbladder. A map of isoplanatic modes is retrieved and is shown to be strongly correlated with the arrangement of tissues constituting the medium. The corresponding focusing laws yield an ultrasound image with drastically improved contrast and transverse resolution. UMI thus provides a flexible and powerful route towards computational ultrasound.
\end{abstract}

\maketitle

\section{Introduction}

In most ultrasound imaging,  the human body is insonified by a series of incident waves. The medium reflectivity is then estimated by detecting acoustic backscatter from short-scale variations of the acoustic impedance. 
An image (spatial map) of reflectivity is commonly constructed using delay-and-sum beamforming (DAS). In this process, echoes coming from a particular point, or image pixel, are \rev{selected by summing the signals generated by this echo at the aperture, thereby accounting – for each element – for the respective time-of-flight associated with the forward and return travel paths of the ultrasonic wave between the probe and that point.} 
\rev{The resulting signal is allocated  at the corresponding pixel of the image, and the procedure repeated for each pixel.} 
The time-of-flight between any incident wave and focal point is calculated with the assumption that the medium is homogeneous with a constant speed-of-sound. However, in human tissue, long-scale fluctuations of the acoustic impedance can 
invalidate this assumption~\cite{hinkelman1997measurements}. 
\la{The resulting incorrectly calculated times-of-flight (also called focusing laws) can lead to aberration of the associated image, meaning that resolution and contrast are strongly degraded.} 
\la{While adaptive focusing methods have been developed to deal with this issue, they rely on the hypothesis that \rev{the speed-of-sound variations occur only in a thin screen at the probe aperture.}}
However, this assumption is simply incorrect \al{in soft tissues}~\cite{Dahl2005} \la{such as fat, skin and muscle, in which the order of magnitude of acoustic impedance fluctuations is around $5\%$~\cite{duck1990acoustic}. This causes higher-order aberrations which} 
are only invariant over small regions\la{, often referred to as \textit{isoplanatic patches.}} 
To tackle this issue, recent studies~\cite{bendjador2020svd, chau2019locally} extract an aberration law for each image voxel by probing the correlation of the \lc{time-delayed} 
echoes coming from adjacent focal spots. The aberration laws are estimated either in the time domain~\cite{montaldo2011time,jaeger2015full,osmanski2012aberration, bendjador2020svd} or in the Fourier domain~\cite{robert2008green,chau2019locally}, for different insonification sequences (focused beams~\cite{montaldo2011time,osmanski2012aberration}, single-transducer insonification~\cite{chau2019locally} or plane wave illumination~\cite{bendjador2020svd}). 
\la{In all of these techniques, a focusing law is estimated in either the receive~\cite{robert2008green, montaldo2011time} or transmit~\cite{Jaeger2015SosMap,chau2019locally} \wi{mode}, but this law is then used to compensate for aberrations \wi{in both reflection and transmission}. }
However, spatial reciprocity between input and output is 
\la{only valid if the emission and detection of waves \wi{are} performed in the same basis; in other words, 
the distortion undergone by a wavefront travelling to and from a particular point is only the same if the wave has interacted with the same heterogeneities in both directions.} 
If this condition is not fulfilled, applying the same aberration phase law in transmit and receive \wi{modes} may improve the image quality \la{to some degree, but will not be optimal.} 

\wil{To \la{obtain} 
optimized focusing laws both in transmit and receive \al{modes}, 
\la{these} two steps 
\la{must} be \alex{considered separately}. Recent studies \cite{lambert2020reflection,Blondel2018,Badon2016} have 
\la{shown how to} \alex{decouple} the location of the transmit and receive focal spots to build \alex{a} focused reflection \al{(FR)} matrix \wi{$\mathbf{R}$}. \la{Containing} 
the medium responses between virtual sources and virtual sensors located within the medium\la{, this matrix is the foundation} 
of \al{ultrasound matrix imaging (UMI)}\la{. Firstly, a focusing criterion and a background rate can be built from the FR matrix, which allows the mapping of  
the focusing quality \rev{and contrast} 
\la{over all pixels of} the ultrasound image\alex{~\cite{lambert_ieee}} in both speckle and specular regimes~\cite{lambert2020reflection}. Secondly,} 
\rev{a} distortion matrix \al{$\mathbf{D}$} 
\la{can be} built from the FR matrix for a local correction of high-order aberrations; \la{this concept was first presented in} 
optical imaging~\cite{Badon2019}, then in ultrasound~\cite{lambert2020distortion}\la{,} and 
\la{most} recently in seismology~\cite{Touma2020}. \la{Whereas} 
$\mathbf{R}$ holds the wave-fronts which are reflected from the medium, 
$\mathbf{D}$ contains the deviations from an ideal reflected wavefront which would be obtained in the absence of heterogeneities. It has been shown that, for specular reflectors~\cite{Badon2019}, \alex{in sparse media~\cite{Touma2020}} 
\la{and} in the speckle regime \rev{for a multi-layered speed-of-sound distribution}~\cite{lambert2020distortion}, a \rev{singular value decomposition (SVD)} of $\mathbf{D}$ yields a one-to-one association between each isoplanatic patch in the focal plane and the corresponding wavefront distortion in the far-field. \lc{This information then enables a correction of aberrations over multiple isoplanatic patches.}} 

\rev{\lc{For laterally-varying aberrations}, the $\mathbf{D}$-matrix should be analyzed locally and investigated over limited spatial windows~\cite{lambert2020distortion}. \rev{In this paper, we generalize this approach to cope with \textit{in-vivo} applications \rev{in which} multiple scattering and high-order aberrations can greatly reduce the size of isoplanatic patches.}} 
\rev{In that respect, the \al{FR} matrix will play a pivotal role. First, it enables the application of an adaptive confocal filter to: (\textit{i}) reduce the detrimental influence of multiple scattering and/or noise; (\textit{ii}) ensure a local isoplanaticity in order to converge towards a satisfying aberration law. Second, \lc{working with} the \al{FR} matrix 
\lc{allows easy projection of the input or output wave-fields into different bases (far-field, \la{transducer} plane, \la{some} intermediate surface\la{,} \textit{etc.}) for optimal aberration correction.} In contrast with previous studies that compensate aberrations from a single (transducer or plane wave) basis~\cite{robert2008green,Jaeger2015SosMap,chau2019locally,bendjador2020svd}, we will show that alternating correction bases is particularly relevant for spatially-distributed aberrations. 
\lc{In each basis, local aberration phase laws can be estimated} by adjusting the field-of-view covered by each local $\mathbf{D}$-matrix with each isoplanatic patch. \lc{By theoretically modelling the SVD process, we show how to optimize} 
the size of the targeted isoplanatic patch and the numerical confocal pinhole. The whole process can then be iterated by gradually reducing the size of \la{the} isoplanatic patches, thereby compensating \la{for} more and more complex aberrations throughout the iteration process.} 

\rev{\lc{The $\mathbf{D}$-matrix formalism was first introduced in the context of ultrasound imaging with simple proof-of-concept experiments ~\cite{lambert2020distortion} in which the points described above (confocal filter, projection in complementary correction bases, theoretical modelling of the SVD process) were not tackled. Here, we show that these important steps make UMI much more robust for \lc{challenging} in-vivo applications.} 
\rev{The overall method is validated by means of \revv{numerical simulations} that mimic aberrations through the abdominal wall.} As an experimental proof-of-concept, we apply UMI to the complex case of \textit{in-vivo} imaging of a \rev{gallbladder}. A set of optimized focusing laws is obtained for each point of the medium\la{, enabling} 
(\textit{i}) a mapping of the isoplanatic \rev{modes} in the field-of-view\la{, revealing} 
the arrangement of the different tissues in the medium
\la{, and} (\textit{ii}) 
\la{the calculation of an ultrasound image with} optimal contrast and 
\rev{enhanced} resolution over \rev{a major part of the} field-of-view. The drastic improvement compared to the conventional ultrasound image is quantified by means of the focusing $F-$factor \rev{and the incoherent background rate} introduced in the first paper of the series~\cite{lambert_ieee}.}

\section{The focused reflection matrix}
\label{par:focusing}

\subsection{Experimental procedure} 
\rev{The experiment was performed on a healthy volunteer (this study is in conformation with the declaration of Helsinki). The probe was placed in a subcostal position in order to image the gallbladder along its short axis. In this configuration, the gallbladder appears as a circular ring.} \la{The experimental \rev{procedure}} consists in recording the reflection matrix $\mathbf{R}$ using a standard plane wave sequence\cite{Montaldo2009}.  The acquisition was performed using a medical ultrafast ultrasound scanner (Aixplorer Mach-30, Supersonic Imagine, Aix-en-Provence, France) driving a \rev{$2-10$} MHz linear transducer array containing \rev{$N=192$} transducers with a pitch $p=0.2$ \rev{mm} (\rev{SL10-2}, Supersonic Imagine). 
The ultrasound sequence 
\la{consisted} in transmitting $101$ steering angles spanning from $-25^{o}$ to $25^{o}$ \rev{with steps of $\delta \theta=0.5^{o}$}, \la{calculated assuming a constant 
speed of sound \rev{$c_0=1540$ m/s}}. \rev{This choice 
\lc{was} made by optimizing the spatially-averaged focusing factor $F$ obtained for a set of speed-of-sound \lc{hypotheses ranging from $c_0=1450-1600$~m/s}
~\cite{lambert2020reflection,lambert_ieee}}. The emitted signal was a sinusoidal burst of three half periods of the central frequency \rev{$f_c=5$}~MHz\la{, with pulse repetition frequency $1000$ Hz}. For each excitation, the back-scattered signal was recorded by 
\la{all probe elements} 
over time 
$\Delta t = 80$ $\mu$\alex{s} \la{with a} 
sampling frequency $f_s=40$ MHz. 
\la{Mathematically, we write this set of acoustic responses as} $\mathbf{R}_{u\theta}(t)\equiv [R(u\out,\tin,t)]$, where $u\out$ defines the \alex{coordinate} of the \alex{receiving} transducer, $\tin$ \alex{the angle of incidence} and $t$ the time-of-flight. 
\la{Subscripts `in' and `out' denote propagation in the forward and backward directions, respectively.} 
Note that the coefficients of $\mathbf{R}_{u\theta}$ should be complex\la{, as they} 
contain the amplitude and phase of the medium response. 
\la{If the responses $\mathbf{R_{u\theta}}(t)$ are not complex modulated RF signal\wi{s}, then the corresponding analytic signals should be considered.}

\subsection{Computing the focused reflection matrix}

In \la{conventional} ultrasound imaging, the reflectivity of a medium at a given point is estimated by 
(\textit{i}) focusing a wave on this point, thereby 
\la{creating} a virtual source
\la{, and} (\textit{ii}) \la{coherently} summing 
the echoes coming from 
\la{that} same point, thus synthesizing a virtual \rev{receiver} at 
\la{that} location. In \al{UMI}, 
\la{this} focusing operation is performed in post-processing\la{,} and the input/output focusing points, $\rin$/$\rout$, are decoupled  \al{[Fig.~\ref{fig:F}(a)]}. This is the principle of the broadband \la{focused reflection} 
matrix 
\la{$\overline{R}_\mathbf{rr} = \overline{R}(\rin,\rout)$} 
containing the responses 
between virtual sources and sensors 
\la{located throughout the medium}. \rev{Note that the concept of virtual transducers is here mainly didactic and that, of course, 
\lc{a virtual transducer does} not act as \lc{a} real source or sink of energy. Moreover, they are 
\lc{highly} directive
\lc{; the directivity pattern points downwards for a virtual source and upwards for a virtual receiver.}} 

In the first article of 
\la{this} series~\cite{lambert_ieee}, the 
\la{FR} matrix was built 
\la{using} a beamforming \alex{process} in the temporal Fourier domain. 
\la{Here, \wi{we show that this matrix \rev{of complex coefficients} can \lc{also} be directly} \lc{(and more quickly)} computed in the time domain \rev{from the recorded IQ signals} via conventional DAS beamforming:} 
\begin{multline}
	\overline{R}(\rout,\rin) = \sum_{\theta\in, u\out} A(u\out, \theta\in,\rin, \rout) \\
	R(u\out,\theta\in, \rev{\tau\in(\theta\in,\rin) + \tau\out(u\out,\rout)}),
	\label{eq:R_rr__time}
\end{multline}
\la{where} $\tau\in$ and $\tau\out$ 
\la{are} the transmit and receive focusing laws such that
\begin{subequations}
    \label{eq:timeOfFlight}
    \begin{align}
&	\tau\in(\theta\in, \rin) =  [ x\in\sin(\theta\in)+z\in\cos(\theta\in)]/\al{c_0} ,\mbox{ and} 
	\\
&	\tau\out(u\out,\rout) = \sqrt{|x\out -u\out|^2 + z\out^2}/\al{c_0}\al{.} 
    \end{align}
\end{subequations}
$A$ \la{is} an apodization factor that 
\la{limits the extent of} the receive synthetic aperture
\la{,} and \alex{($x\in,z\in$) and ($x\out,z\out$) are the coordinates of the input and output focusing points $\rin$ and $\rout$\la{, respectively}. In this paper, we will restrict our study to the $x-$projection of $\overline{\mathbf{R}}_{\wi{\mathbf{rr}}}$, 
\la{written} $\overline{\mathbf{R}}_{xx}(z)$, in which only the responses between virtual transducers located at the same depth are considered ($z=z\in=z\out$).}


\subsection{Manifestation of aberrations and multiple scattering}
\begin{figure}
    \centering
    \includegraphics[width=0.9\columnwidth]{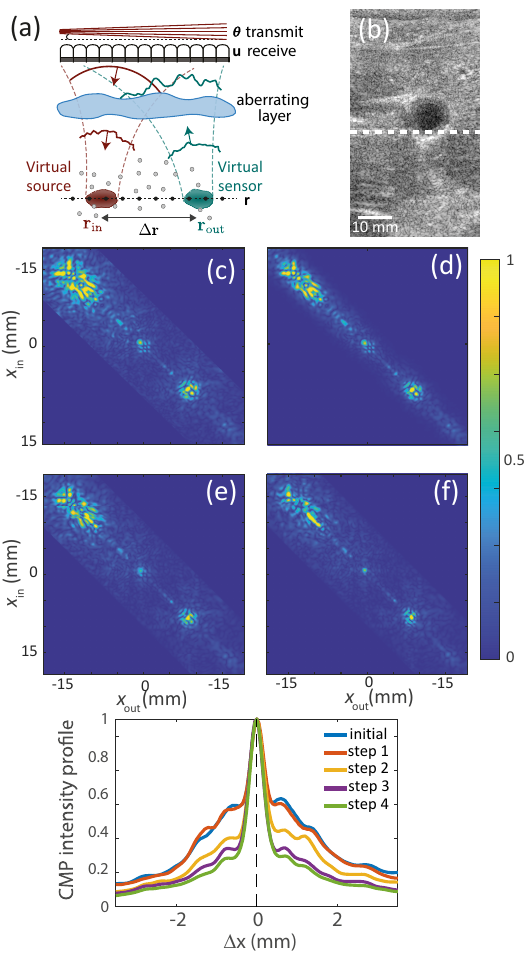}
    \caption{\al{Focused reflection matrix.} (a) \al{UMI} consists in splitting the location\la{s} of the transmit ($\rin$) and receive ($\rout$) \la{focusing} points. \rev{(b) Conventional ultrasound image of the gallbladder.} \al{\rev{(c)}-(f)} \alex{Evolution} of 
    \la{$\mathbf{R}_{xx}(z)$ at depth \rev{$z = 39$~mm} \rev{(white dashed line in (b)}}
    during the aberration correction process. \la{$\overline{\mathbf{R}}_{xx}$ is shown \al{(b)} prior to correction, \al{(c)} after the application of the adaptive confocal filter \eqref{eq:confocalFilter}, \al{(d,e)} after the first and fourth iteration of the aberration correction process, respectively.} \rev{(g) Evolution of the corresponding CMP intensity profiles $I(\mathbf{r},\Delta x)$ spatially averaged over the area $\mathcal{B}$ \rev{[Fig. \ref{fig:Full_results}(c)]}.}}
    \label{fig:F}
\end{figure}

\wil{Each row of $\overline{\mathbf{R}}_{xx}(z)$ 
\la{corresponds to the situation in which waves have been focused at $\rin=(x_\textrm{in},z)$} in transmission, and virtual detectors at $\rout=(x_\textrm{out},z)$ record the resulting spatial wave spreading across the focal plane} 
\al{[Fig.~\ref{fig:F}(a)]}. \al{Fig.~\ref{fig:F}(b)} shows 
\la{$\overline{\mathbf{R}}_{xx}(z)$} at 
\rev{$z=39$~mm}. \alex{Note that the coefficients $\overline{R}(x\out, x\in,z)$ associated with a transverse distance $|x\out - x\in|$ larger than a superior bound $\Delta x_{\mathrm{max}}\sim \lambda_{\mathrm{max}}/(2\delta \theta)$ are not displayed because of spatial aliasing~\cite{lambert_ieee}. \rev{The angle increment $\delta \theta$ could have been decreased to remove this spatial aliasing; however, this would be at the cost of a longer measurement time which is critical for in-vivo imaging since the medium is moving.}}
\alex{
\la{In the single scattering approximation, the} FR matrix \rev{coefficients} can be expressed \la{theoretically} as follows~\cite{lambert_ieee}:
\begin{equation}
\label{Rrr_intermsofh_eq2}
\overline{R}(x\out,x\in,z)=\int d x  H_\text{out}(x,x\out,z)  \gamma(x,z) H_\text{in}(x,x\in,z),
\end{equation}
where $\gamma(x,z)$ is the medium 
reflectivity. ${H}\in(x,x\in,z)$ and ${H}\out(x,x\out,z)$ are the transmit and receive 
\la{PSFs}, \textit{i.e.} the spatial amplitude distribution of the input and output focal spots at depth $z$.} In \lc{the} absence of aberration, their spatial exten\lc{t} 
$\overline{\delta x}_0(\mathbf{r}_\text{in/out})$ scales as $ \lambda /[2 \sin \beta (\mathbf{r}_\text{in/out}) ]$, where $\beta(\mathbf{r})$ \lc{is} the maximum angle of 
\lc{illumination} (in transmit mode) or collection (in receive mode) by the array from the associated focal point $\mathbf{r}$. 

\begin{figure}
    \centering
    \includegraphics[width=0.8\columnwidth]{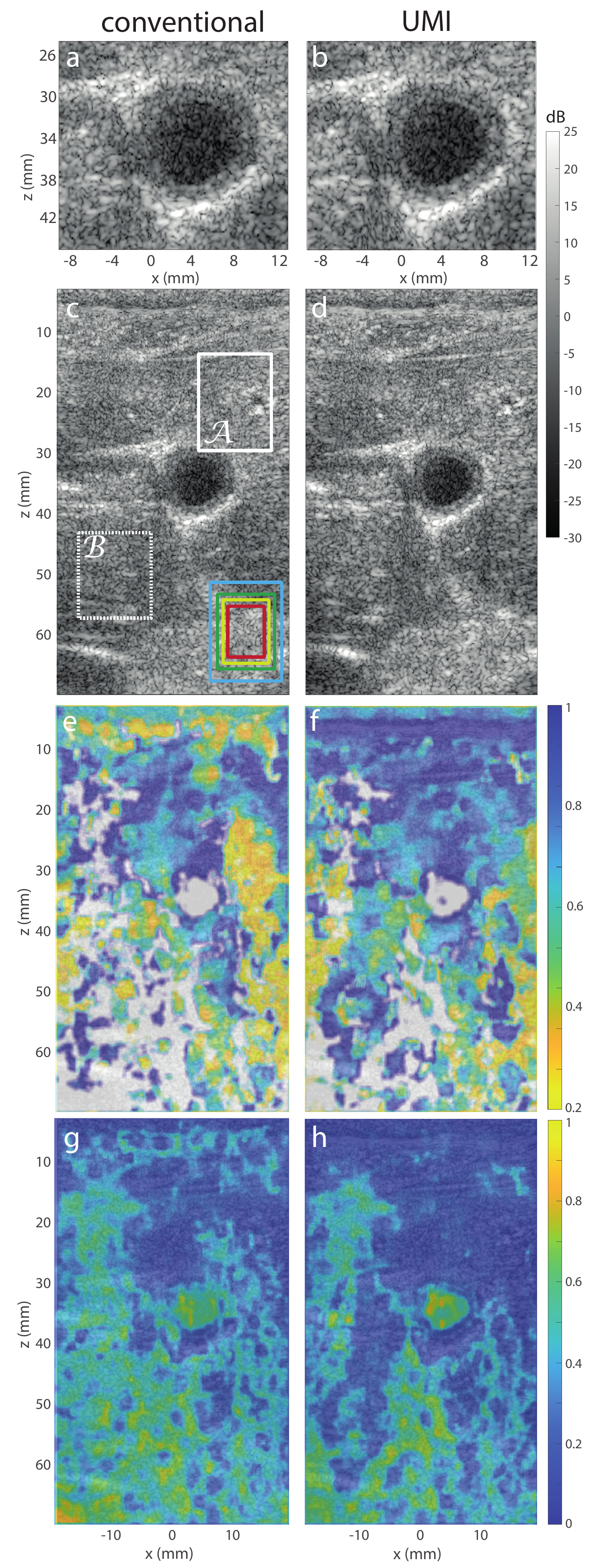}
    \caption{\rev{Results of the aberration correction process applied to 
    {\textit{in-vivo} imaging} of the gallbladder. (a,b) Zooms on the gallbladder of the initial and corrected UMI images displayed in (c,d), respectively.} \rev{(c) Conventional \rev{dynamically focused (\textit{i.e} confocal)} image. \rev{The {continous} white square $\mathcal{A}$ shows the area where the correlation matrix displayed in Fig.~\ref{fig:correlationMatrix}(a) is computed.} The {dashed} white square \rev{$\mathcal{B}$} defines the area used 
   {to estimate} the average {CMP} intensity profiles {shown} in Fig.~\ref{fig:F}\rev{(g)}. The four {solid,} colored rectangles correspond to the spatial windows {$W_{\Delta \mathbf{r}}$} used {in the four steps} of the correction {process} (from the largest to smallest, {see Table.\ref{tab:parameters}}). (b) Corrected \rev{dynamically focused} {UMI} image. (e,f) Focusing criterion superimposed {on}to the (e) {conventional} and (f) {corrected} images}. \rev{(g,h) Incoherent background rate superimposed {on}to the (g) {conventional} and (h) {corrected} images.}}
    \label{fig:Full_results}
\end{figure}

\rev{The intensity distribution along the diagonal of $\overline{\mathbf{R}}_{xx}(z)$ yields a line of the \rev{confocal} image \al{$\mathcal{I}$} that would be obtained by plane wave synthetic beamforming~\cite{Montaldo2009}: 
\begin{equation}
\label{imcalc}
\mathcal{I}(\mathbf{r})\equiv \left|\al{\overline{R}}\left(\mathbf{r},\mathbf{r} \right)\right|^2.
\end{equation}
The experimental ultrasound image is displayed in Figs.~\ref{fig:Full_results}(a) and (b). 
\lc{In Fig.~\ref{fig:Full_results}(a), the blurred appearance of the gallbladder internal wall indicates the presence of aberrations and/or multiple scattering. We now quantify these detrimental effects} by investigating the off-diagonal coefficients of $\mathbf{R}_{xx}$.}

In the accompanying paper~\cite{lambert_ieee}, the intensity of the antidiagonals of $\overline{\mathbf{R}}_{xx}(z)$, referred to as the common mid-point \al{(CMP)} intensity profile, \al{is} shown to give access to the local input-output incoherent PSF, $|H\in|^2 \stackrel{\al{\Delta x}}{\circledast} |H\out|^2 $  \rev{in the speckle regime} \al{(the symbol $\circledast$ here stands for a convolution product)}. \al{Fig.}~\ref{fig:F}\al{(g)} shows the 
\la{CMP} intensity profile \rev{in the area $\mathcal{B}$ displayed in Fig.~\ref{fig:Full_results}(c)}. 
A confocal peak can be observed, originating from single scattering, which sits on \lc{a} wider incoherent background \rev{resulting from high-order aberrations, multiple scattering and experimental noise}. \rev{By applying the fitting procedure developed in the accompanying paper~\cite{lambert_ieee}, maps of the focusing factor $F(\mathbf{r})$ and of the incoherent background rate $\alpha (\mathbf{r})$ 
\lc{were} extracted from local CMP intensity profiles (see Figs.~\ref{fig:Full_results}(e) and (g), respectively).}  
Blue areas in Fig.~\ref{fig:Full_results}(e)
\al[$F(\mathbf{r})\sim 1)$] indicate high reliability; the \rev{ultrasound} 
image accurately describes the medium reflectivity (see Fig.~\ref{fig:Full_results}\al{(a,c)}).  
\rev{Green and yellow} areas 
\la{[$F(\mathbf{r})<0.5$] indicate aberrated areas of the image} . 
Gray areas correspond to the situation where the \rev{incoherent background is too large ($\alpha>0.75$) to obtain a reliable estimation of the focusing factor $F$.} \la{This occurs} when the intensity level of the backscattered signal generated by the region of interest is \rev{threefold} lower than \alex{the multiple scattering and/or electronic noise contributions} (see yellow regions in Fig.~\ref{fig:Full_results}(g)). 
\lc{Causes of this low intensity level could be weak reflectivity of the medium in the probed region, and/or strong fluctuations of the speed-of-sound upstream \lc{of (at shallower depths than)} the focal plane, which would decrease the relative single scattering level close to the diagonal.} \rev{
\lc{Overall}, the ultrasound image displayed in Fig.~\ref{fig:Full_results}(c) suffers from: (\textit{i}) a focusing quality that is far from 
\lc{ideal} in many 
\lc{areas} (especially at shallow depths and on the right part of the image); (\textit{ii}) a weak contrast due to a predominant incoherent background (especially at large depths and on the left part of the image)}.   \la{Thus, prior to performing aberration correction, it is important to remove as much \rev{incoherent} background as possible from the FR matrix.}

\subsection{Filtering multiple scattering and noise}\label{par:confocal_filter}

The \rev{incoherent background can be partially suppressed} using an adaptive confocal filter~\cite{Blondel2018, Badon2016}. This process consists in weighting the coefficients $\overline{R}(x\in, x\out,z)$ of the \al{FR} matrix as a function of the distance $ | x\out - x\in | $ between the virtual transducers, such that:
\begin{equation}
    R'(  x\out, x\in  ) = \overline{R}( x\out, x\in ) \exp{ \left [ - \frac{\mid  x\out - x\in  \mid^2}{2 l_c^2(\mathbf{r})} \right ]}.
    \label{eq:confocalFilter}
\end{equation}
This filter has a Gaussian shape, with a width $l_{c}(\mathbf{r})$ that needs to be carefully set. If $l_c(\mathbf{r})$ is too large, the multiply-scattered echoes will \la{prevent} a correct estimation of the aberration phase law. If $l_c(\mathbf{r})$ is too small, the filter then acts as an apodization function that will smooth out the resulting aberration phase law. \rev{To be efficient, $l_c(\mathbf{r})$ \rev{is automatically chosen to scale with} $\overline{\delta x}_0(\mathbf{r})$ (hence adaptive): $l_c(\mathbf{r})= n \wi{\overline{\delta x}_{0}}(\mathbf{r})$, with $n$ a coefficient whose value is reported in \al{Table}.~\ref{tab:parameters}. The choice of this value will be justified in Sec.~\ref{par:isoplanatic}.}  
\begin{table}
    \renewcommand{\arraystretch}{1.4}
    \centering
            \caption{\label{tab:parameters} \rev{Parameters used for the UMI process}.}
    \begin{tabular}{|c|c|c|c|c|c|}
        \hline
        \multicolumn{2}{|c|}{Correction steps} & 1 & 2 & 3 & 4 \\
                 \hline
           \multicolumn{2}{|c|}{Transmit basis}   & $k\in$ & $u\in$ & $k\in$ & $u\in$ \\
        \hline
          \multicolumn{2}{|c|}{Receive basis} & $u\out$ & $k\out$ & $u\out$ & $k\out$ \\
        \hline
        \multirow{3}{1cm}{\begin{center} exp.\end{center}}
        & $l_c/\wi{\overline{\delta x}_{0}}$ & $7.5 $ & $10 $ & $12 $ & $12$ \\
        \cline{2-6}
       & $\Delta x$ (mm) & 12 & 9.6 & 7.7 & 6.2\\
        \cline{2-6}
      & $\Delta z$ (mm) & 16 & 12 & 10.2 & 8.1 \\
         \hline
      \multirow{2}{1cm}{\begin{center} num.\end{center}}
   & $\Delta x$ (mm) & 10 & 7 & 5 & 3\\
         \cline{2-6}
   &  $\Delta z$ (mm) & 10 & 7 & 5 & 3 \\
         \hline
    \end{tabular}
\end{table}

\al{Figs.~\ref{fig:F}(b,c) show the original and filtered FR matrices}, $\overline{\mathbf{R}}_{xx}$ and \al{${\mathbf{R}}'_{xx}$}, respectively, 
\lc{for} \rev{$z=39$~mm}. \la{It can be seen that the adaptive confocal filter has removed part of the \rev{incoherent background}.} However, \al{${\mathbf{R}}'_{xx}$} still contains a residual multiple scattering \rev{and/or noise} component \la{which \rev{also pollutes} matrix coefficients very close to the diagonal.}  
Note that this filter has no impact on the raw ultrasound image since the confocal signals are unaffected. However, it constitutes a necessary step for \rev{an unbiased estimation} of the aberration law.

\section{Matrix correction of aberrations}

\begin{figure}[htbp]
  \centering
   \includegraphics[width=0.9\columnwidth]{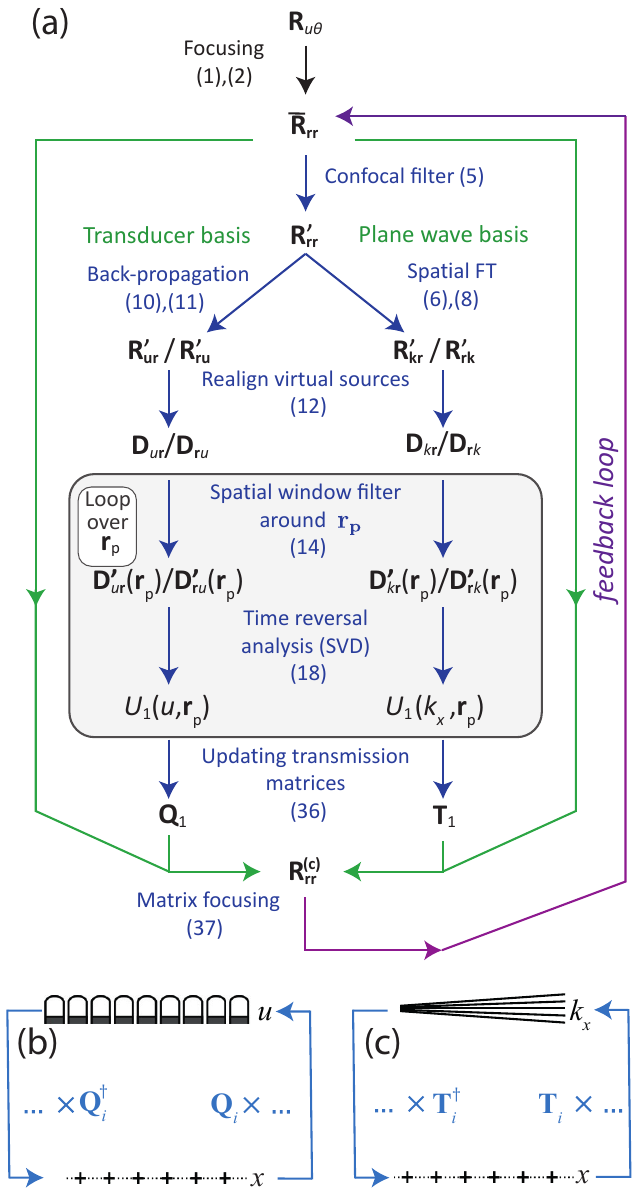}
   \caption{(a) Workflow of the \rev{various} steps of the aberration correction process.  The blue lines refer to the estimation process of the aberration phase laws while the green lines refer to the aberration correction procedure. \wi{The purple line depicts the iteration of the UMI process.} The gray rectangle symbolizes a loop over the spatial window centered on $\rp$. \rev{(b,c) Schematic representation for the projection between the transducer/far-field and focused bases.}}
   \label{fig:schema__process_abCorr1}
\end{figure}

The FR matrix ${\mathbf{R}}'_{xx}(z)$ is now used to implement the distortion matrix concept~\cite{lambert2020distortion}. \la{In this section, we will show how to}  estimate and correct for aberrations successively 
in the transmit and receive modes, both from the far-field and the 
\la{surface of the transducer array}. 
\la{We describe all of} the technical steps of the aberration correction process: (\textit{i}) the projection of the FR matrix at output or input in\la{to} a correction basis (here either the Fourier or transducer basis) in order to investigate the reflected or incident wave-front associated with each virtual source or transducer, respectively
\la{,} (\textit{ii}) the realignment of the \al{transmitted or} reflected wave-fronts to form the distortion matrix $\mathbf{D}$
\la{,}  (\textit{iii}) the truncation of $\mathbf{D}$ into overlapping isoplanatic patches
\la{,}  (\textit{iv}) the singular value decomposition of $\mathbf{D}$ or of its normalized correlation matrix to extract an aberration phase law 
\la{for each isoplanatic patch, and (\textit{v}) the application of the focusing law and back-projection of the reflection matrix into the focused basis. All of these steps are repeated by exchanging input and output bases,} as well as the correction basis. \rev{Fig. \ref{fig:schema__process_abCorr1} shows a workflow that sums up the different steps of the UMI procedure.} 
The 
process \rev{is} 
\la{then} be iterated while gradually reducing the size of isoplanatic patches in order to address higher order aberrations.  


\subsection{Projection of the reflection matrix into 
a dual basis}
\label{par:basis_change}

In adaptive focusing, the aberrating layer is often modeled as a phase screen. For an optimal correction, \la{the} ultrasonic data should \al{be} back-propagated to the plane containing 
\la{this} aberrating layer\la{; indeed, \rev{from} this plane,} 
the aberration is spatially-invariant. By applying the phase conjugate of the aberration phase law, aberration 
 can be fully compensated \la{for} at any point of the medium. However, in real life, speed-of-sound inhomogeneities are distributed over the 
\la{entire} medium and aberration 
can take place everywhere. 
\la{To treat this case, the strategy here} is to back-propagate ultrasound data in\la{to} several planes from which the aberration phase law should be estimated and then compensated. 
\la{\al{The optimal correction plane is the one that maximizes the size of isoplanatic patches.} For multi-layered media, the \al{Fourier} plane is the most adequate since plane waves are the propagation invariants in this geometry. For aberrations induced by \wi{superficial veins or skin nodules}, 
the probe plane is a good choice. In this paper, the aberration correction will be performed in these two planes as they coincide also to the emission and reception bases used to record the reflection matrix. However, note that, in practice, other correction planes can be chosen according to the imaging problem.} 
\\

\subsubsection{Projection into the Fourier basis}

To project the reflected wave-field into \lc{the} \rev{Fourier plane}, a spatial Fourier transform should be applied to the \la{output} of \al{$\mathbf{R}'_{xx}(z)$} \rev{(Fig.~\ref{fig:schema__process_abCorr1}(c))}:
\begin{equation}
    \al{\mathbf{R}'_{kx}}(z) = \mathbf{T}_0 \times \al{\mathbf{R}'_{xx}}(z).
\label{eq:Rkz_projection}
\end{equation}
where $\mathbf{T_0}$ is the Fourier transform operator
\begin{equation}
\label{T0}
    T_{0}(k_x,x) = \exp{(i k_{x} x)}/\sqrt{N_k},
\end{equation}
\la{$k_x$ the transverse wave number\rev{, and $N_k$ the dimension of $\mathbf{T_0}$}. \rev{On the one hand, the support $\Delta k$ of the wave-field in the Fourier plane is fixed by the resolution of the focal plane: $\Delta k \sim 1 / \overline{\delta x}_0$. On the other hand, its sampling $\delta k$ is the inverse of the field-of-view : $\delta k \sim 1/(N\overline{\delta x}_0)$. This choice ensures a conservation of the information between the focused and Fourier bases, such that $\mathbf{T_0}\mathbf{T}_0^{\dag}=\mathbb{I}$, where $\mathbb{I}$ is the identity matrix and $\dag$ stands for the transpose conjugate operation. } 
} \al{$\mathbf{{R}}'_{kx}(z) \al{\equiv [R(k\out,x\in,z)]}$} contains the set of aberrated wavefronts \rev{in the spatial Fourier domain} generated by each virtual source $\rin$. \al{Fig.}~\ref{fig:matrix_D}(c) shows the phase of \al{$\mathbf{{R}}'_{kx}(z)$} 
\la{at} \rev{$z=60$} mm. Using the central frequency $f_c$ as a reference frequency, the transverse wave number \al{$k\out$} can be associated with a plane wave of angle \al{${\theta}\out$}, such that $\al{k\out} =k_c \sin(\al{\theta\out})$, with \la{$k_c=2\pi f_{c}/\al{c_0}$.} 
Expressing the far-field projection as a plane wave decomposition is useful to define the boundaries of this basis [white dashed lines \al{in Fig.}~\ref{fig:matrix_D}(a)]\la{; the maximum transverse wave number is $k_\textrm{max} \sim  k_c \sin [\beta (\mathbf{r})]$ [see \al{Fig.}~\ref{fig:matrix_D}\rev{(k)}].} 

\la{The matrix $\mathbf{{R}}'_{kx}(z)$ will be used to tackle aberrations in the receive plane wave basis. To do the same in the transmit basis, a reciprocal projection to that of \eqref{eq:Rkz_projection}} 
can be performed at the input of \al{$\mathbf{{R}}'_{xx}(z)$}:
\begin{equation}
\label{Rxk}
    \al{\mathbf{R}'_{xk}}(z) = \al{\mathbf{{R}}'_{xx}}(z) \times \mathbf{T}_0^{\top}.
\end{equation}
\wi{where the symbol $\top$ stands for matrix transpose}. The coefficients \al{$R'(x\out,k\in,z)$} \la{correspond} 
to the wave-field probed by the virtual transducer at $\rout$ if a plane wave of transverse wave number \al{$k\in$} illuminates the medium.  This matrix will be used to investigate 
aberrations in the transmit plane wave basis.
\\

\subsubsection{Projection into the transducer basis}

\la{The strategy to treat aberrations in the transducer plane is similar to that described above.} 
Free-space transmission matrix, $\mathbf{Q}_{0}=[Q_0(\mathbf{r},\mathbf{u})]$, \al{is} defined between the focused and 
transducer bases at 
\la{central} frequency $f_c$:
\begin{equation}
    \mathbf{Q}_{0}= \mathbf{T}_{0}^{\dag} \times \left( \alex{\mathbf{P} }\circ \mathbf{T}_{0} \right),
    \label{eq:Q_0}
\end{equation}
\alex{where the symbol $\circ$ stands for 
\la{a} Hadamard product and $\mathbf{P}=[P(k_x,z)]$ is the plane wave propagator at the central frequency: $
    P(k_x,z)=e^{i \sqrt{k_c^2-k_x^2}z}$.} \rev{The transmission matrix $  \mathbf{Q}_{0}$ can be considered here at the single central frequency because of the time gating operation performed in \eqref{eq:R_rr__time}. }\rev{If evanescent waves are neglected, the sampling $\delta u$ in the transducer basis can be fixed to $\lambda_c/2$, with $\lambda_c=c_0/f_c$ the wavelength at the central frequency $f_c$, such that $\mathbf{Q}_0\mathbf{Q}_0^{\dag}=\mathbb{I}$.}
\alex{The operator $\mathbf{Q}_0$ can be given a physical interpretation by reading the terms of \eqref{eq:Q_0} from right to left: (i) a spatial Fourier transform using the operator $\mathbf{T}_{0}$ to project the wave-field in the plane wave basis; (ii) the plane wave propagation modeled by the propagator $\mathbf{P}$ between the focal and 
\la{transducer} planes over a distance $z$; (iii) an inverse Fourier transformation $\mathbf{T}_{0}^{\dag}$ that 
\la{projects} the wave-field in\la{to} the transducer basis. \revv{Note that the coefficients of $\mathbf{Q}_0$ actually correspond to the $z-$derivative of the free space Green's functions between transducers and focusing points at the central frequency $f_c$.}} 

 \alex{Using this operator $\mathbf{Q}$, the matrix \al{$\mathbf{R}'_{xx}$} can be projected in\la{to} the transducer basis \rev{(Fig.~\ref{fig:schema__process_abCorr1}(b))} either at input,}
\begin{equation}
    \label{eq:Rru__correction}
    \al{\mathbf{R}'_{xu}}(z) = \al{\mathbf{R}'_{xx}}(z) \times \mathbf{Q_0}^{\top}
\end{equation}
or output,
\begin{equation}
       \label{eq:R_ur__correction}
    \al{\mathbf{R}'_{ux}}(z) = \mathbf{Q_0} \times \al{\mathbf{R}'_{xx}(z).}
\end{equation}
Each column of $\al{\mathbf{R}'_{xu}}(z)\alex{=[R(x\out,u\in,z)]}$ 
\la{holds} the wave-field received by the virtual transducer at $\rout$ for an incident wave-field emitted from a transducer at $u\in$. Reciprocally, each row of $\al{\mathbf{R}'_{ux}}(z)=[R(u\out,x\in,z)]$ contains the wave-front recorded by the transducers for a virtual source in the focal plane at $\rin$ \rev{(Fig.~\ref{fig:distortion_virtual}(a))}. \al{Fig.~\ref{fig:matrix_D}\rev{(i)}} shows the phase of $\al{\mathbf{R}'_{ux}}(z)$ obtained at \rev{$z=60$~mm.} \al{At 
\la{this relatively large depth,} the spatial extension $\Delta u$ of the reflected wave-field in the transducer basis coincides with the physical aperture $A$ of the array used to collect the echoes coming from a depth $z$. \la{In contrast, {Fig.~\ref{fig:matrix_D}\rev{(k)}} demonstrates that for shallower depths} 
$z<A \tan [\beta (\mathbf{r})] /2$, $\Delta u$ is limited by the numerical aperture of the probe 
such that $\Delta u \sim 2 z \tan [ \beta_\textrm{max}]$.} 

\subsubsection{Discussion}
\begin{figure*}
    \centering
    \includegraphics[width=1\textwidth]{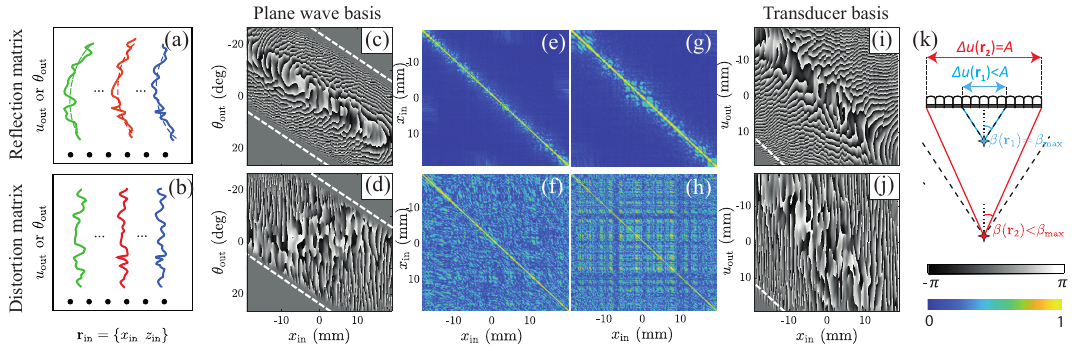}
    \caption{\al{Revealing the spatial correlations between reflected wave-fields. \rev{By subtracting (a) the phase of the dual reflection matrix (sketched by continous lines) by its geometrical counterpart (dashed line), (b) a distortion matrix is computed and reveals the long-range spatial correlations exhibited by the reflected wave-fronts over each isoplanatic patch. (c,d) Phase of the dual reflection matrix, $\mathbf{R}'_{kx}$, and distortion matrix, $\mathbf{D}_{kx}$, in the plane wave basis. (e,f) Corresponding spatial correlation matrices of their phase, $\mathbf{C}^{(R)}_{xx}$ and $\mathbf{C}^{(D)}_{xx}$, respectively.  {(g,h)} Phase of the dual reflection matrix, $\mathbf{R}'_{ux}$, and distortion matrix, $\mathbf{D}_{ux}$, in the transducer basis. (i,j) Corresponding spatial correlation matrices of their phase, $\mathbf{C}^{(R)}_{xx}$ and $\mathbf{C}^{(D)}_{xx}$, respectively. The matrices shown in panels {(c)-(j)} correspond to a depth {of} $z=60$~mm. (k) The white dashed lines in panels (c)-(d) and (g)-(h) account for the finite angular extent $\beta(\mathbf{r})$ and spatial support $\Delta u (\mathbf{r})$ of the reflected waves in the plane wave and transducer bases, respectively.}}}
    \label{fig:matrix_D}
\end{figure*}

We might expect to observe correlations between the columns of matrices \al{$\mathbf{R}'_{kx}(z)$} and \al{$\mathbf{R}'_{ux}(z)$} displayed in Figs.~\ref{fig:matrix_D}\rev{(c)-(i)}. 
\la{Because neighboring} virtual sources $\rin$ belong \textit{a priori} to the same isoplanatic patch, 
the associated wave-fronts \la{observed} in the transducer plane or in the far-field should thus be, in principle, strongly correlated since they travel through the same area of the aberrating layer. \rev{Such correlations can be revealed by the spatial correlation matrix, $\mathbf{C}_{xx}^{(R)}=\exp(j\mbox{arg} \lbrace{\mathbf{R}^{'\dag}_{vx}} \rbrace )\times \exp(j\mbox{arg} \lbrace{\mathbf{R}{'}_{vx}}\rbrace)$ (with $v=k$ or u). However, whether it be in the Fourier (Fig.~\ref{fig:matrix_D}(e)) or transducer bases (Fig.~\ref{fig:matrix_D}(g)), $\mathbf{C}_{xx}^{(R)}$ exhibits a diagonal feature characteristic of uncorrelated wave fronts between the columns of the reflection matrices.} In the following, we show how to reveal 
\rev{the} hidden correlations in the reflection matrix by introducing the distortion matrix. 
\la{We will \al{consider} 
mostly} the transducer basis\la{,} as the far-field case has already been 
\la{explored} in a previous work~\cite{lambert2020distortion}. 

\subsection{The distortion matrix}

\alex{To reveal the isoplanaticity of the reflected wave-field,} each aberrated wave-front contained in the reflection matrix \al{$\mathbf{R}'_{ux}(z)$} \wi{[Fig.~\ref{fig:matrix_D}\rev{(a)}]} should be decomposed into two components: (\textit{i}) a geometric component described by $\mathbf{Q}_{0}(z)$, which contains the ideal wave-front induced by the virtual source $\rin$ that would be obtained in the homogeneous medium used to model the wave propagation \rev{[see dashed lines in Fig.~\ref{fig:matrix_D}{(a)}]}; (\textit{ii}) a distorted component due to the mismatch between the propagation model and reality [Fig.~\ref{fig:matrix_D}\rev{(b)}]. A key idea is to isolate the latter contribution by subtracting, \rev{from the phase of the experimentally measured wave-field, its ideal counterpart}. \alex{Mathematically, this operation can be done by means of an 
Hadamard product between \al{$\mathbf{R}'_{ux}(z)$} and $\mathbf{Q}^*_{0}(z)$}:
\begin{equation}
    \alex{\mathbf{D}_{ux}(z) =  \al{\mathbf{R}'_{ux}(z)} \circ \mathbf{Q}_{0}^{*}(z).}
    \label{eq:Distortion_matrix}
\end{equation}
\wi{where the symbol $*$ stands for phase conjugate.} {We call} $\wi{\mathbf{D}_{u\mathbf{r}}}=\mathbf{D}_{ux}(z)=[D({u}\out, \{x\in,z\})]$ 
the \textit{distortion matrix}. The distortion matrix connects any input focal point $\rin$ to the distorted component of the reflected wavefield in the transducer basis.

\rev{To clarify the physical meaning of the distortion matrix, its coefficients can be expressed} {in the \al{Fraunhoffer} approximation as~\cite{Goodman1996}}:
\begin{equation}
\label{eqD2}
    D({u}\out, \{x\in,z\}) =
    \sum_{\delta x} \al{R'(x\in+\delta x,x\in,z)} e^{i \frac{k_c}{2z}  u\out \delta x} ,
\end{equation}
with $\delta x=x\out-x\in$. Mathematically, each \wi{column} of \wi{$\mathbf{D}_{u\mathbf{r}}$} is the \alex{Fourier} transform of the focused wave-field re-centered around each focusing point $\rin$. \wi{$\mathbf{D}_{u\mathbf{r}}$} can thus be seen as a \al{dual} reflection matrix for different realizations of virtual sources, all shifted at the origin of the focal plane ($x_{in} = 0$, \rev{see Fig.~\ref{fig:distortion_virtual}(b))}. The co-location of the virtual sources at the same point 
\la{is the reason why \rev{the phase laws along each column of} $\mathbf{D}_{ux}(z)$ [Figs.~\ref{fig:matrix_D}\rev{(j)}] \rev{are} much more correlated than those of \al{$\mathbf{R}'_{ux}(z)$}} [Fig.~\ref{fig:matrix_D}\rev{(g)}]. \rev{This observation is quantitatively confirmed by the corresponding correlation matrix, $\mathbf{C}_{xx}^{(D)}=\exp(j\mbox{arg} \lbrace{\mathbf{D}^{\dag}_{ux}} \rbrace )\times \exp(j\mbox{arg} \lbrace{\mathbf{D}_{ux}}\rbrace)$, displayed in Fig.~\ref{fig:matrix_D}\rev{(h)}. It shows much larger off-diagonal correlation coefficients than the original reflection matrix  $\mathbf{R}'_{ux}$ (Fig.~\ref{fig:matrix_D}\rev{(g)}).} 

\rev{The matrix operation in \eqref{eq:Distortion_matrix} is equivalent to the 
\lc{classic} guide star principle of adaptive focusing~\cite{flax1988phase}. When considering the reflection matrix $\mathbf{R}'_{ux}$ between focused and transducer bases in the transmit and receive modes, respectively, the focused transmission is the synthetically generated guide star. The phase conjugate of the propagation matrix $\mathbf{Q}_{0}$ is the delay compensation one would conventionally use prior to determining the aberration law at the aperture data. The distortion matrix $\mathbf{D}_{u \mathbf{r}}$ is the conventional delay-corrected aperture data generated by each synthetic guide star.}

\alex{\wi{Note that equivalent distortion matrices, \al{$\mathbf{D}_{\mathbf{r}u}$, $\mathbf{D}_{k \mathbf{r}}$ and $\mathbf{D}_{\mathbf{r} k}$} can be built from the other reflection matrices previously defined\la{:} 
\al{$\mathbf{R}'_{xu}(z)$, $\mathbf{R}'_{k x}(z)$ and $\mathbf{R}'_{x k}(z)$}.} For \al{$\mathbf{D}_{\mathbf{r}u}$}, the same reasoning as above can be used by exchanging input and output. 
\la{The} far-field distortion matrices, \wi{\al{$\mathbf{D}_{k \mathbf{r}}$ and $\mathbf{D}_{\mathbf{r} k}$}} 
\la{have already been investigated in a previous work~\cite{lambert2020distortion}}. The comparison between the phase of \al{$\mathbf{R}'_{k x}\wi{(z)}$ [Fig.~\ref{fig:matrix_D}\rev{(c)}] and $\mathbf{D}_{k x}\wi{(z)}$} [Fig.~\ref{fig:matrix_D}\rev{(f)}] highlights the high 
\la{degree of correlation} of the distorted \rev{wave-fronts} in the far-field\la{, resulting from} 
the virtual shift of all the input focal spots 
\la{to} the origin. \rev{Again this is unambiguously quantified when comparing the corresponding correlation matrix $C_{xx}^{(D)}$ (Fig.~\ref{fig:matrix_D}\rev{(f)}) with its original counterpart  $C_{xx}^{(R)}$ (Fig.~\ref{fig:matrix_D}\rev{(e)}).} }

\subsection{Local distortion matrices}
\label{par:local}
\la{We have shown that virtual sources which} 
belong to the same isoplanatic patch 
should give rise to strongly correlated distorted wave-fronts\la{,} 
\la{even if the reflectivity of the medium is random} [see Figs. \ref{fig:matrix_D}\wi{(e,f)}]. To correct for multiple isoplanatic patches in the field-of-view, recent works~\cite{lambert2020distortion, Badon2019} show that the distortion matrix can be analyzed \alex{over the whole field-of-view}. Its \alex{effective} rank is then equal to the number of isoplanatic \rev{modes} \alex{contained in this field-of-view}, while its \alex{singular} vectors yield the corresponding aberration phase laws. The \alex{proof}-of-concept of this fundamental result 
\la{was first} demonstrated in optics for specular reflectors \cite{Badon2019}, then in ultrasound speckle for multi-layered media~\cite{lambert2020distortion} \wi{and lately in seismic imaging for sparse media~\cite{Touma2020}.}

\la{Here, we investigate the case of ultrasound \textit{in vivo} imaging, in which}  
fluctuations of the speed-of-sound occur both in the lateral and axial directions. 
\la{This means that the spatial distribution of aberration effects can become more complex. \rev{In Fig.~\ref{fig:Full_results}\al{(e)},} strong fluctuations of the $F$-map} illustrate the \rev{relatively small size of isoplanatic patches that can only be induced by a complex distribution of the speed-of-sound in the medium.} \la{Such  complexity implies that} any point in the medium will be associated with \la{its own} distinct aberration phase law. \la{To construct an image in these conditions, the transmission matrix connecting the correction and focused bases should therefore be constructed to include all of these phase laws -- an extremely difficult task. Here, this problem will be tackled using an analysis of a \textit{local} distortion matrix.}  The idea is to take advantage of the local \al{isoplanicity} of the aberration phase law around each focusing point. 

To begin, we divide the field-of-illumination into overlapping regions that are defined by their central midpoint $\rp$ and their spatial extension $\Delta \mathbf{r} = \{\Delta x,\Delta z\}$. All of the distorted components associated with focusing points $\rin$ located within each region are extracted and stored in a local distortion matrix $\mathbf{D}'_{u \mathbf{r}}(\rp)$:
\begin{equation}
\label{eq:window2}
    D'({u}\out, \rin, \rp) = D(u\out, \rin) ~ W_{\Delta \mathbf{r}}(\rin - \rp),
\end{equation}
where $W_{\Delta \mathbf{r}}(\mathbf{r}) = 1$ for $|x|<\Delta x$ and $|z|<\Delta z$, and zero otherwise. Ideally, each sub-distortion matrix should contain a set of focusing points $\rin$ belonging to the same isoplanatic patch. In reality, the isoplanicity condition is never completely fulfilled. A delicate compromise thus has to be made on the size $\Delta \mathbf{r}$ of the window function: it must be small enough to approach the isoplanatic condition, but large enough to encompass a sufficient number of independent realizations of disorder~\cite{lambert2020distortion}. This last point 
\la{is} discussed in \rev{Sec.~\ref{par:aberrationLaw}}.


\subsection{Isoplanicity} \label{par:isoplanatic}

\rev{The isoplanatic hypothesis can be ensured by means of the adaptive confocal filter described in Sec.~\ref{par:confocal_filter}. The parameter $l_c$ can actually control the extension of the aberrated PSF in the filtered matrix $\mathbf{R'}_{xx}$. If we model the spatially-distributed aberrations at each depth $z$ by a thin aberrating layer located at $z/2$, $l_c$ should scale as $l_c \sim \lambda z/(2l_p)$, 
\lc{where $l_p$ is} the coherence length of the aberrator. 
\lc{As $l_p$ also governs the isoplanatic length in the focal plane~\cite{Mertz2015}, the isoplanatic hypothesis will be fulfilled if $\Delta x \sim l_p$.} 
\lc{This, combined with the fact that $\overline{\delta x}_0 \sim \lambda z/(Np)$ in the far-field, gives $l_c \sim n \overline{\delta x}_0$, where $n \sim  Np/({2 \Delta x})$; this relation dictated the choice of extension $l_c$ of the adaptive confocal filter} in Table \ref{tab:parameters}. }

\alex{\rev{T}he isoplanatic condition can now be assumed to be fulfilled over each region of size $\Delta \mathbf{r}$.} \alex{This hypothesis implies that the PSFs ${H\in}$ and ${H\out}$ are invariant by translation in each region. \rev{This leads us to define a local spatially-invariant
PSF $\rev{H^{(l)}_\textrm{in/out}}$ around each central midpoint $\rp$ such that:}} \la{$H_\textrm{in/out}(x',x,z) = \rev{H^{(l)}_\textrm{in/out}}(x'- x,z,\rp)$. 
 Injecting \eqref{Rrr_intermsofh_eq2} into \eqref{eqD2} leads to the following expression for the $\mathbf{D}$-matrix coefficients:
\begin{multline}
   \al{D'}({u}\out, \rin , \rp ) =   \\
  \rev{{\tilde{H}}^{(l)}\out}(u\out,\rp) \int dx\gamma(x+x\in,z) \rev{H^{(l)}\in}(x,\rp)  e^{i \frac{k_c}{2z}  u\out x} \la{.}
  \label{Dalex}
\end{multline}
The physical meaning of this last equation is the following: Around each point $\rp$, the aberrations can be modelled by a transmittance  $\rev{\tilde{H}^{(l)}\out}(u\out,\rp)$. This transmittance is 
the Fourier transform of the output PSF $\rev{H^{(l)}\out}(x,\rp)$:
\begin{equation}
\rev{{\tilde{H}}^{(l)}\out}(u\out,\rp)= \sum_{x} \rev{H^{(l)}\out}(x,\rp) e^{-i \frac{k_c}{2z}  u\out x} \la{.}
  \label{eq:aberrationLaw_Uout}
\end{equation}
The aberration matrix \rev{$\mathbf{\tilde{H}}^{(l)}\out$} directly provides the \textit{true} transmission matrix $\mathbf{Q}$ between the transducers and any point $\rp$ of the medium:
\begin{equation}
   \mathbf{Q} =\rev{\mathbf{\tilde{H}}^{(l)}\out} \circ \mathbf{Q}_0 \la{.}
\end{equation}
This transmission matrix $\mathbf{Q}$\al{, or equivalently the aberration matrix \rev{$\mathbf{\tilde{H}}^{(l)}\out$,}} \wi{are} the holy grail for ultrasound imaging since \wi{their} phase conjugate directly provides the focusing laws that need to be applied on each transducer to optimally focus on each point $\rp$ of the medium. }


\subsection{Singular value decomposition} \label{par:SVD}
  To extract the aberration \alex{phase} law $\rev{{\tilde{H}}^{(l)}\out}(u\out,\rp)$ from each \alex{local} distortion matrix, we can notice from \eqref{Dalex} that each line of $\al{\mathbf{D}'_{u\wi{\mathbf{r}}}}(\rp)$ is the product between $\rev{{\tilde{H}}^{(l)}\out}(u\out,\rp)$ and a random term associated with each virtual source. 
  To unscramble the deterministic term  ${\tilde{H}}\out(u\out,\rp)$ from the random virtual source term in \eqref{Dalex}, the singular value decomposition (SVD) of $\mathbf{D}'_{u\mathbf{r}}(\rp)$ 
 can be applied. 
  The SVD consists in writing $\mathbf{D}'_{u\wi{\mathbf{r}}}(\rp)$ as
  \begin{equation}
\label{eq:svd}
\mathbf{D}'_{{u}\wi{\mathbf{r}}}(\rp) =\mathbf{U}(\rp) \times \mathbf{\Sigma}(\rp) \times \mathbf{V}^{\dag}(\rp),
\end{equation}
\wi{where the symbol $\dag$ stands for transpose conjugate.} $\mathbf{\Sigma}$ is a diagonal matrix containing the singular value\la{s} $\sigma_i(\rp)$ in descending order: $\sigma_1>\sigma_2>..>\sigma_N$. $\mathbf{U}(\rp)$ and $\mathbf{V}(\rp)$ are unitary matrices that contain the orthonormal set of output and input eigenvectors, $\mathbf{U}_i(\rp)=[U_i({u}\out, \rp)]$ and $\mathbf{V}_i(\rp)=[V_i(\mathbf{r}_\textrm{in}, \rp)]$. 
\la{The physical meaning of this SVD can be intuitively understood by considering the asymptotic} case of a point-like input focusing beam [$\rev{H^{(l)}\in}(x,\rp)=\delta(x)$]. In this ideal case, \eqref{Dalex} becomes $ D({u}\out, \rin , \rp ) =  
\rev{{\tilde{H}}^{(l)}\out}(u\out,\rp) \gamma(\rin)$. 
\la{Comparison with \eqref{eq:svd}} shows that 
$\mathbf{D}'_{{u}\wi{\mathbf{r}}}(\rp)$ is then of rank $1$ \la{-- the} 
first output singular vector $\mathbf{U}_1(\rp)$ yields the aberration transmittance $\rev{\mathbf{\tilde{H}}^{(l)}\out}(\rp)$ while the first input eigenvector $\mathbf{V}_1(\rp)$ directly provides the medium reflectivity. 

However, in reality, the input PSF $H\in$ is of course far from being point-like. \al{As we will show below, the spectrum of $\mathbf{D}'_{u\wi{\mathbf{r}}}(\rp)$ displays a continuum of 
\la{singular} values \rev{(Fig.~\ref{fig:correlationMatrix}(b)). The support of $\mathbf{U}_1(\rp)$ is limited by a correlation length $\delta u_1$ related to the spatial extent $\overline{\delta x}\in$ of the input PSF\lc{,} and $\mathbf{V}_1(\rp)$ only provides a low-resolution image of the medium \rev{reflectivity since its spatial frequency content is also limited by $\delta u_1$}.} Interestingly,} \al{the normalized first output singular vector, $\hat{U}_1(u\out,\rp)=U_1(u\out,\rp)/|{U}_1(u\out,\rp)|$,} can \la{still} constitute \rev{an} estimator of $\rev{\tilde{\mathbf{H}}^{(l)}\out}(\rp)$ \la{in this case}. \rev{However, several effects will induce a bias on this estimator. }\rev{First, the medium should ideally exhibit a random reflectivity; any local 
\lc{structure} such as a plane reflector can alter the estimation of the aberration phase law.} \rev{Second, in this speckle regime, the medium's random reflectivity should be smoothed out \rev{enough} by considering a spatial window $W_{\Delta r}$ containing a large number $N\in$ of independent input 
\la{focal} points \rev{$\mathbf{r}\in$}}. \rev{
\lc{Finally, just like correlation techniques which rely} on the synthetic guide star principle~\cite{flax1988phase}, the spatial extension $\overline{\delta x}_\textrm{in}$ of the transmitted focal spot degrades the quality of 
\lc{the estimator $\hat{U}_1(u\out,\rp)$}. However, unlike correlation techniques, the SVD bias is of a different nature and can be mitigated for a sufficiently large number $N\in$ of disorder realizations. These aspects are now investigated in detail by considering the correlation matrix of $\mathbf{D}'_{{u}\wi{\mathbf{r}}}(\rp)$ in the transducer basis.}     
\begin{figure*}
    \centering
    \includegraphics[width=1\textwidth]{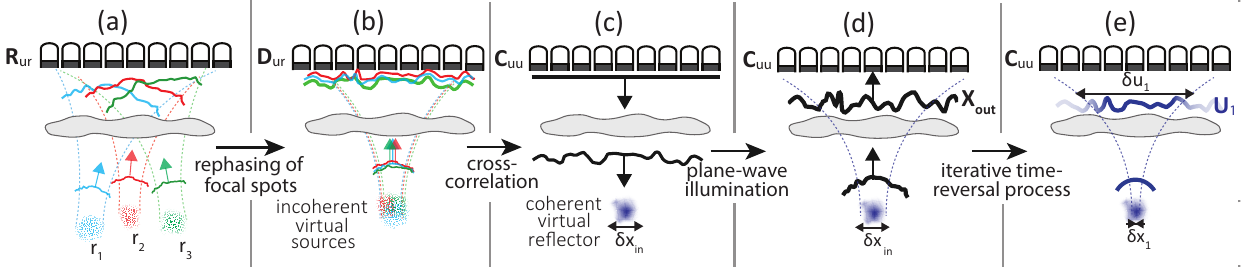}
    \caption{\rev{Sketch of the time reversal analysis of the distortion matrix. (a) Each \al{column} of the reflection matrix $\mathbf{R}_{u \mathbf{r}}$ corresponds to the reflected wavefield induced by the associated virtual source $\mathbf{r_{in}}$. (b) By removing the geometrical curvature of each reflected wavefront \eqref{eq:Distortion_matrix}, the resulting distortion matrix extracts the aberrated component of those wavefronts. From an other point of view, all the wavefronts are realigned as if they were generated by input focal spots that are virtually shifted at the origin \eqref{eqD2}. (C) The correlation matrix {$\mathbf{C}_{uu}$} of {$\mathbf{D}_{u \mathbf{r}}$} mimics the time reversal operator applied to a virtual reflector that results from a {coherent} average of all the shifted input focal spot{s}. \rev{Applying an incident plane wave to {$\mathbf{C}_{uu}$} is the first step of the time reversal process. (d) The resulting wave-field $X\out$ is the estimator of the aberration phase law extracted by conventional correlation techniques. (e) This estimator can be improved by iterating the time reversal process, or equivalently performing the eigenvalue decomposition of $\mathbf{C}_{uu}$ \eqref{eq:evd}. The phase conjugate of its first eigenvector $\mathbf{U}_1$} then yields {the phase law} to focus on the brightest part of this coherent reflector.}}
    \label{fig:distortion_virtual}
\end{figure*}

\subsection{{Correlation matrix}}
\label{par:aberrationLaw}
\rev{To study the SVD of the distortion matrix $\mathbf{D}_{{u}\mathbf{r}}'(\rp)$ in the transducer basis, the correlation matrix $\mathbf{C}_{{u}{u}}(\rp)$ is needed:
  \begin{equation}
    \mathbf{C}_{{u}{u}}(\rp) = {N\in^{-1}}\mathbf{D}_{{u}\mathbf{r}}'(\rp) \times \mathbf{D}_{{u}\mathbf{r}}'^{\dagger}(\rp),
    \label{eq:correlation_matrix}
\end{equation}
with $N\in$ the number of virtual sources \alex{contained in each spatial window $\Delta \mathbf{r}$}. \rev{The SVD of \al{$\mathbf{D}'(\rp)$} is indeed equivalent to the eigenvalue decomposition of $\mathbf{C}_{{u}{u}}(\rp)$:
\begin{equation}
\label{eq:evd}
\mathbf{C}_{{u}{u}}(\rp) =\mathbf{U}(\rp) \times \mathbf{\Sigma}^2(\rp) \times \mathbf{U}^{\dag}(\rp),
\end{equation}
or, in terms of matrix coefficients,
\begin{equation}
\label{eq:evd_coef}
\mathbf{C}_{{u}{u}}(\rp) = \sum_i \sigma_i^2(\rp) U_i(u\out,\rp) U_i^*(u\out,\rp).
\end{equation}
The eigenvalues $\sigma_i^2$ of $\mathbf{C}_{{u}{u}}(\rp)$ are the square of the singular values of $\mathbf{D}'_{{u}r}(\rp)$. The eigenvectors $\mathbf{U}_i(\rp)$ of $\mathbf{C}_{{u}{u}}(\rp)$ are the output singular vectors of $\mathbf{D}'_{{u}r}(\rp)$. The study of $\mathbf{C}_{{u}{u}}(\rp)$ should thus lead to the prediction of 
\lc{singular vectors $\mathbf{U}_i(\rp)$}. }}

\rev{The coefficients of $\mathbf{C}_{{u}{u}}$ can be seen as an average over $\rin$ of the spatial correlation of each distorted wave-field:
\begin{equation}
    C(u,{u}', \rp) = \frac{1}{N\in} \sum_{\rin} D'({u},\rin, \rp) ~ D'^{*}({u}',\rin, \rp).
    \label{eq:correlation_matrix_coefficients}
\end{equation}
 \rev{In \lc{the} absence of the adaptive confocal filter defined in \eqref{eq:confocalFilter}, each element $C(u\out,u\out^\prime,\rp)$ of $\mathbf{C}_{{u}{u}}(\rp)$ would correspond the correlation coefficient between the time-delayed complex (IQ) signals recorded at the ballistic time by the receiving elements, $u\out$ and $u\out^\prime$, averaged over the set of input focusing points $\mathbf{r}_\textrm{in}$ contained in the spatial window centered on $\rp$~\cite{chau2019locally}.} 
 $\mathbf{C}_{uu}$ can be decomposed as the sum of a covariance matrix $\left \langle \mathbf{C}_{uu}\right \rangle $ and a perturbation term $\mathbf{N} $:
\begin{equation}
\label{C}
   \mathbf{C}_{uu}= \left \langle \mathbf{C}_{uu}\right \rangle+  \mathbf{N}.
\end{equation}
$\mathbf{C}_{uu}$ will converge towards $\left \langle \mathbf{C}_{uu}\right \rangle $ 
 if the perturbation term \al{$\mathbf{N}$} tends towards zero. In fact, the intensity of $\mathbf{N}$ scales as the inverse of the number $N\in$ of resolution cells in each sub-region~\cite{robert2008green,Robert_thesis}. In the following, we will thus assume a convergence of $\mathbf{C}$ towards its covariance matrix $\left \langle \mathbf{C}\right \rangle$ due to disorder self-averaging. }

\rev{The covariance matrix can be derived analytically in the speckle regime for which the medium reflectivity $\gamma(\mathbf{r})$ is assumed to be random, meaning that $\langle \gamma(\mathbf{r}) \gamma^*(\mathbf{r}')\rangle =  \langle|\gamma|^2\rangle \delta(\mathbf{r}-\mathbf{r}')$. Under this assumption, injecting (13) into \eqref{eq:correlation_matrix_coefficients} leads to:
\begin{multline}
    C({u}\out, {u}\out',\rp) \propto \langle|\gamma|^2\rangle \rev{\tilde{H}\out}({u}\out,\rp) \rev{\tilde{H}^{(l)}\out}^*({u}'\out, \rp) \\
    \rev{\left[\tilde{H}^{(l)}\in \ast
    \tilde{H}^{(l)}\in \right]}\left({u}\out - {u}'\out,\rp \right),
    \label{eq:<Cuu>}
\end{multline}
where the symbol \al{$\ast$} stands for a correlation product. \alex{The} correlation term, \rev{$\tilde{H}^{(l)}\in \al{\ast} \tilde{H}^{(l)}\in$}, results from the Fourier transform of the input PSF intensity \rev{$\left |H^{(l)}\in \right |^2$}. Equation \eqref{eq:<Cuu>} is reminiscent of the Van Cittert-Zernike theorem for an aberrating layer~\cite{Mallart1991}. This theorem states that the spatial correlation of a random wavefield generated by an incoherent source is equal to the Fourier transform of the intensity distribution of this source (here the input aberrated focal spots). }

\begin{figure}
    \centering
    \includegraphics[width=1\columnwidth]{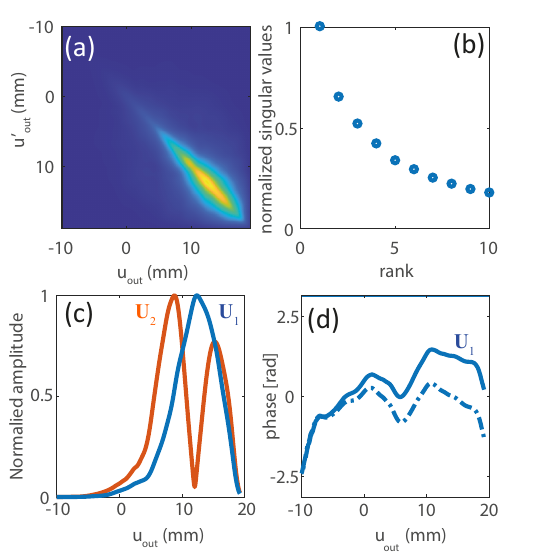}
    \caption{\rev{Extraction of the aberration phase law from the correlation matrix in the output transducer basis computed over the area $\mathcal{A}$ in Fig.~\ref{fig:Full_results}(c). (a) Modulus of $\mathbf{C}_{uu}$. (b) Ten first normalized eigenvalues of the correlation matrices $\mathbf{C}_{uu}$. (c) Modulus of the two first eigenvectors of $\mathbf{C}_{uu}$, $\mathbf{U}_1$ (blue line) and $\mathbf{U}_2$ (red line).  (d) The aberration phase law $\hat{\mathbf{U}}_1$, before (continuous blue line) and after the linear phase ramp correction (dashed blue line).}}
    \label{fig:correlationMatrix}
\end{figure}

\rev{Let us write the eigenvalue decomposition of the positive semidefinite correlation kernel, $\tilde{\mathbf{H}}^{(l)}\in \ast \tilde{\mathbf{H}}^{(l)}\in =\left [ \tilde{H}^{(l)}\in \ast
    \tilde{H}^{(l)}\in \left({u}\out - {u}'\out  \right) \right ]$,
    \begin{equation}
    \label{evd_kernel}
  \tilde{\mathbf{H}}^{(l)}\in \ast \tilde{\mathbf{H}}^{(l)}\in =  \mathbf{W} \times \mathbf{L} \times \mathbf{W}^\dag 
    \end{equation}
    or, in terms of matrix coefficients,
     \begin{equation}
    \label{evd_kernel_coeff}
  \tilde{{H}}^{(l)}\in \ast \tilde{{H}}^{(l)}\in \left({u}\out - {u}'\out\right) = \sum_{i=1}^N l_i \mathbf{W}_i \left({u}\out \right) \mathbf{W}_i^* \left({u}'\out \right)
    \end{equation}   
   with $\mathbf{L}$ a diagonal matrix whose real coefficients $l_i$ are the eigenvalues of  $\tilde{\mathbf{H}}^{(l)}\in \ast \tilde{\mathbf{H}}^{(l)}\in$ ranged in decreasing order. $\mathbf{W}$ is a unitary matrix whose columns, $\mathbf{W}_i=[W_i(u\out)]$, correspond to the eigenvectors of  $\tilde{\mathbf{H}}^{(l)}\in \al{\ast} \tilde{\mathbf{H}}^{(l)}\in$. \rev{By injecting \eqref{evd_kernel_coeff} into \eqref{eq:<Cuu>}, one obtains the following expression for $\mathbf{C}_{uu}$:
      \begin{equation}
    \label{evd_Cuu}
 \mathbf{C}_{uu} \propto  \left [ \tilde{\mathbf{H}}^{(l)}\out \circ \mathbf{W} \right ] \times \mathbf{L} \times \left [ \mathbf{W}  \circ \tilde{\mathbf{H}}^{(l)}\out \right ]^\dag
    \end{equation}   
Still under the assumption that the aberrations only induce phase retardation effects ($|\tilde{H}^{(l)}\out({u}\out,\rp)|=1$), the matrix $\tilde{\mathbf{H}}^{(l)}\out \circ \mathbf{W}$ is unitary. Because of the uniqueness of the eigenvalue decomposition, Eqs.~\ref{eq:evd} and \ref{evd_Cuu} show that the eigenvalue distribution of $\mathbf{C}_{uu} $ is dictated by its correlation kernel ($\sigma_i^2 \propto l_i$) and that its eigenvectors $\mathbf{U}_i$ are given by:
\begin{equation}
\label{theory}
    \mathbf{U}_i(\rp) \propto \mathbf{\tilde{H}}^{(l)}\out(\rp) \circ \mathbf{W}_i(\rp).
\end{equation} }}
\lc{Thus, the determination of whether the eigenvectors $ \mathbf{U}_1(\rp) $ can be satisfactory estimators of $\tilde{\mathbf{H}}^{(l)}\out$ rests on the properties of  $\mathbf{W}_i(\rp)$, the eigenvectors of $\tilde{\mathbf{H}}^{(l)}\in \ast \tilde{\mathbf{H}}^{(l)}\in$.}
\rev{$\mathbf{W}_i(\rp)$ can be derived by solving a second order Fredholm equation with \lc{a} Hermitian kernel ~\cite{robert2009prolate,ghanem2003stochastic}. An analytical solution can be found for certain analytical form of the correlation function $\tilde{H}^{(l)}\in \ast \tilde{H}^{(l)}\in$}
\lc{: in the absence of aberration [$\tilde{H}^{(l)}\in(u\in) = 1$],} 
\rev{
{$\tilde{H}^{(l)}\in \ast \tilde{H}^{(l)}\in$} should be equal to a triangle function that spreads over the whole correlation matrix {~\cite{robert2008green}}.  
In presence of \lc{aberration}
, a significant drop of the correlation width $\delta u_{in}$ of $\tilde{H}^{(l)}\in \ast \tilde{H}^{(l)}\in$ is expected. $\delta u_{in}$ is actually inversely proportional to the spatial extent $\overline{\delta x}\in$ of the input PSF $H\in$: $\delta u_{in}\sim \lambda z/\overline{\delta x}\in$~\cite{Mallart1994}.  Fig.~\ref{fig:correlationMatrix}{(a)} illustrates that fact by showing the {modulus} of the correlation matrix $\mathbf{C}_{{u}{u}}(\rp)$ computed over the area $\mathcal{A}$ in Fig.~\ref{fig:Full_results}(c). If we assume that the aberrations only induce phase retardation effects ($|\tilde{H}^{(l)}\out({u}\out,\rp)|=1$), the modulus of $\mathbf{C}_{uu}$ is actually a direct estimator of {$\tilde{\mathbf{H}}^{(l)}\in {\ast} \tilde{\mathbf{H}}^{(l)}\in$}. As shown by Fig.~\ref{fig:correlationMatrix}{(a)}, the correlation function $\tilde{H}^{(l)}\in \ast \tilde{H}^{(l)}\in$ is far from having a triangular shape and it decreases rapidly with the distance $|{u}\out - {u}'\out|$. For such a bounded correlation function, the effective rank of $\mathbf{C}_{uu}$ is shown to scale as the number of resolution cells contained in the input PSF $H\in$~\cite{robert2009prolate}:
\begin{equation}
\label{rank}
    M_\delta \sim (\Delta u /\delta u\in) \sim (\overline{\delta x}\in /\delta x_0)\lc{.}
\end{equation}
\rev{The shape of the corresponding eigenvectors $\mathbf{W}_i(\rp)$ depends on the exact form of the correlation function, or equivalently on the shape of the virtual scatterer. For instance, a flat reflector (sinc correlation function) 
\lc{gives} 3D prolate spheroidal eigenfunctions\cite{robert2009prolate}; a cylindrical object leads to Hermite-Gaussian eigenmodes\cite{aubry2006gaussian}. 
As the correlation function $\tilde{H}^{(l)}\in \al{\ast} \tilde{H}^{(l)}\in$ is, in first approximation, real and positive (\textit{i.e} associated with a symmetric PSF envelope $|\tilde{H}^{(l)}\in|$), a general trend is that the first eigenvector $\mathbf{W}_1(\rp)$ shows a nearly constant phase. The phase of the first eigenvector $\mathbf{U}_1(\rp)$ is \rev{then} a direct estimator of $\mathbf{\tilde{H}}^{(l)}\out(\rp)$ [blue \rev{continuous} line in Fig. \ref{fig:correlationMatrix}\rev{(d)}].} \rev{In practice, $\mathbf{W}_1(\rp)$ can exhibit a linear phase ramp due to the asymmetry of the input PSF envelope $\left |{H}^{(l)}\in \right |$. 
\lc{This} results in an additional phase ramp in $\mathbf{U}_1(\rp)$ compared to the targeted aberration phase law $\tilde{H}^{(l)}\out({u}\out,\rp)$. We will provide in Sec.~\ref{par:physical_interpretation} a physical interpretation of this problem and a method to circumvent it.}}

\rev{A more critical issue comes from the 
\lc{fact that $\mathbf{W}_1(\rp)$ (and thus $|\mathbf{U}_1(\rp)|$) exhibits} a \rev{bell curve shape whose characteristic width $\delta u_1$ is dictated by the spatial extent of the input PSF: $\delta u_1 \sim \sqrt{\lambda z \left ( 1+ z/\overline{\delta x}\in \right)}\sim z \sqrt{\lambda /\overline{\delta x}\in }$~\cite{aubry2006gaussian}. The higher rank eigenvectors $\mathbf{W}_i(\rp)$ are more complex and exhibit a number of lobes that 
\lc{scale} with their rank $i$.}} \rev{  Fig.~\ref{fig:correlationMatrix}\rev{(c)} show\lc{s} the modulus of the first two eigenvectors of the matrix $\mathbf{C}_{uu}$ 
\lc{shown} in Fig.~\ref{fig:correlationMatrix}(a). We recognize the typical signature of the two first eigenmodes with one and two lobes respectively.}
For aberration correction, it is thus important to consider the normalized vector $\hat{\mathbf{U}}_1=[U_1(u\out)/|U_1(u\out)]$ that only implies a phase shift
\lc{, rather than the original vector \alex{$\mathbf{U}_1$} as is done in Ref.~\cite{bendjador2020svd}.}  \rev{In the latter case, the bounded support of \alex{$\mathbf{U}_1$} 
\lc{will} limit the numerical aperture to $\delta u_1/z$
\lc{, and, for strong aberrations, deeply degrade the resolution of the corrected image.} 
\lc{However, the} shape of $|\mathbf{U}_1|$ implies \rev{that the 
\lc{aberration phase law estimated from $\hat{\mathbf{U}}_1$ is inconsistent on the edges of its support.}
This bias is induced by the perturbation term in \eqref{C} whose variance scales as $\left \langle |N(u\out,u'\out)|^2\right \rangle= \left \langle \left |C(u\out,u\out)\right|^2 \right \rangle /N_\textrm{in}$. 
\lc{Taking a perturbational approach, $\mathbf{U}_1$ can be written as the sum} of its expected value, $\mathbf{W}_1 \circ \tilde{\mathbf{H}}^{(l)}\out$, and a 
\lc{first-order} perturbation \rev{term $\delta \mathbf{U}_1$, with $|\delta U_1(u\out)|^2 \sim M_\delta^2/(N\in N\out)$. This decomposition can be seen as the sum of a constant phasor and a weak random phasor that implies the following phase error on the eigenvector $\mathbf{U}_1$: $\left \langle | \mbox{arg} \lbrace U_1 \rbrace \right|^2 \rangle = |\delta U_1|^2 /|W_1|^2$~\cite{Goodman2000}. Assuming a cylindrical virtual scatterer~\cite{aubry2006gaussian}, $|W_1(u\out)|^2 \sim (M_1/N\out) \exp \left ( - (u\out-u_0)^2 / \delta u_1^2\right )$, with $u_0$ the center of the $\mathbf{U}_1$ support. The phase error can then be expressed as: 
\begin{equation}
\label{randomphasor}
\left \langle | \mbox{arg} \lbrace U_1 (u\out) \rbrace \right |^2 \rangle \sim \frac{M_\delta^2}{N\in M_1} \exp \left ( u\out^2 / \delta u_1^2\right )
\end{equation}}
\rev{with $M_1=(\Delta u/\delta u_1)\sim \sqrt{\lambda \delta x\in}/\overline{\delta x}_0$. The exponential term in \eqref{randomphasor} confirms that the SVD bias mainly occurs on the edges of the array in the transducer basis. A correct estimation of the aberration phase law \rev{\eqref{theory}} over the whole probe aperture is obtained provided that :
\begin{equation}
\label{con}
    N_\textrm{in} > \frac{M_\delta^2}{M_1}\exp (M_1^2) .
\end{equation}}
This condition is quite restrictive and justifies our initial choice for the area $\Delta \mathbf{r}$: $N\in = \Delta x \Delta z/(\overline{\delta x}_0 \delta z_0)\sim 5000$ (see Table~\ref{tab:parameters}) \rev{with $\delta z_0 = c / (2 \Delta f)$, the axial resolution of the ultrasound image}. It also explains why the aberration correction process should \wi{then} be iterated\la{; at} 
each iteration, the focal spot size, $\overline{\delta x}_\textrm{in}$ or $\overline{\delta x}_\textrm{out}$, decreases and the spatial window $\Delta \mathbf{r}$ can be reduced accordingly. \la{In the end, the measurement of} 
aberration matrices $\mathbf{\tilde{H}}^{(l)}\out$ and $\mathbf{\tilde{H}}^{(l)}\in$ 
\la{will thus have} 
excellent spatial resolution.}}

\subsection{{Time reversal picture}} \label{par:physical_interpretation} 

\rev{
\lc{In this section, we describe the SVD process using a time reversal picture. This picture allows for an intuitive physical interpretation of the theoretical 
concepts introduced in the previous section, and highlights} the difference\lc{s between our aberration correction approach and other, more conventional} 
techniques~\cite{flax1988phase,chau2019locally,Jaeger2015SosMap}.
\lc{We} also show how the aforementioned additional phase ramp may arise in the estimated aberration phase law\lc{,} and how to correct for this artifact. }

\rev{
\lc{In the previous section, we showed} that the aberration phase law can be extracted from the SVD of the distortion matrix $\mathbf{D}'_{{u}r}(\rp)$. This operation can be 
\lc{seen} as a fictive time reversal experiment. Expressed in the form of \eqref{eq:<Cuu>}, $\mathbf{C}_{uu}$ is analogous to a reflection matrix $\mathbf{R}$ associated with a single scatterer of reflectivity $|H\in(x)|^2$ [Fig.~\ref{fig:distortion_virtual}(c)].
For such an experimental configuration, it has been shown that an iterative time reversal process converges towards a wavefront that focuses perfectly through the heterogeneous medium onto this scatterer~\cite{Prada1994,Prada1996}. Interestingly, this time-reversal invariant can also be deduced from the \rev{first eigenvector} of the time-reversal operator $\mathbf{R}\mathbf{R}^{\dag}$~\cite{Prada1994,Prada1996,Prada2003}. The same decomposition could thus be applied to $\mathbf{C}_{uu}$ in order to retrieve the wavefront that would perfectly compensate for aberrations and optimally focus on the virtual reflector. This effect is illustrated in Fig.~\ref{fig:distortion_virtual}(e). }

\rev{This time reversal picture illuminates the difference \lc{between the approach in this article, and the} 
correlation techniques that are 
\lc{widely} used for aberration compensation. The latter 
basically build an estimator $X\out$ of the aberration phase law by a simple average of the correlation coefficients~\cite{Silverstein2003}: $X\out(u\out)=N\out^{-1}\sum_{u'\out} C(u\out,u'\out)$. This operation is equivalent to the first iteration of the iterative time reversal procedure when a plane wave is first emitted from the transducers to initiate the process (Fig.~\ref{fig:distortion_virtual}(c)). $X\out$ then corresponds to the wave-field reflected by the virtual scatterer (Fig.~\ref{fig:distortion_virtual}(d). 
\lc{The input PSF intensity $|H^{(l)}\in(x)|^2$ is blurred due to aberration, meaning that the virtual scatterer (or, equivalently, the guide star) has an enlarged spatial extent. An aberration phase law estimated using such a enlarged/distorted guide star will have an inherent bias.} 
Indeed, using \eqref{eq:<Cuu>}, the cross-correlation estimator $X\out$ can be expressed as follows:
\begin{multline}
   X\out(u\out) \propto \tilde{H}^{(l)}\out({u}\out,\rp) \left[\tilde{H}^{(l)}\in \ast
    \tilde{H}^{(l)}\in \circledast \tilde{H}^{(l)}\out \right]\left({u}\out,\rp \right),
    \label{eq:<X>}
\end{multline}
The estimator $X\out$ thus yields the aberration phase law (left term in \eqref{eq:<X>}) modulated by a correlated random phase term (right term in \eqref{eq:<X>}) with a coherence length scaling with $\delta u\in \sim \lambda z /\delta x\in$. The estimator thus exhibits a strong bias due to the input PSF blurring~\cite{flax1988phase}.  }

\rev{This bias can be circumvented by iterating the time reversal process, or equivalently by performing a SVD of the distortion matrix. Indeed, the iteration of the time reversal process tends to maximize the energy back-scattered by the virtual reflector~\cite{varslot2004eigenfunction,aubry2006gaussian,robert2009prolate}. It yields a wave-front $\mathbf{U}_1$ that tends to focus on the center of the virtual reflector over a resolution length $\delta x_1 \sim \lambda z/\delta u_1 \sim  \sqrt{\lambda \delta x\in} <\delta x\in $.  The SVD estimator of the aberration phase law is thus less impacted by the blurring of the input focal spot and the presence of multiple scattering and/or noise than standard correlation techniques based on the guide star principle.}

\rev{However, another issue may occur if the scattering distribution $\left |H^{(l)}\in(x) \right|^2$ is too complex. $\mathbf{U}_1$ then focuses on the brightest spot exhibited by the input PSF $\left |H^{(l)}\in(x) \right|^2$. The phase of $\mathbf{U}_1$ can then exhibit an additional linear phase ramp compared to the \textit{true} aberration phase law~\cite{varslot2004eigenfunction}:
\begin{equation}
\label{phase_ramp}
    \hat{{U}}_1(u\out)=\tilde{H}\out(u\out)e^{-i \frac{k_c}{2z}  u\out x_0}
\end{equation}
where $x_0$ corresponds to the lateral shift of the corresponding PSF $H_1$, such that
\begin{equation}
\label{PSF1}
{H}_1(x)= \sum_{u\out} \hat{U}_1(u\out) e^{i \frac{k_c}{2z}  u\out x} ={H}_1(x-x_0).
\end{equation}
 If no effort is made to remove this shift, each selected area defined by the spatial window function $W_{\mathbf{\Delta r}}$ could suffer from arbitrary lateral shifts $x_0$ compared to the original image. This artifact can be suppressed by removing the linear phase ramp in \eqref{phase_ramp}. One way to do 
 \lc{this is to estimate $x_0$ using} the auto-convolution product of the incoherent PSF $|{H}_1|^2$:
 \begin{equation}
\label{PSF2}
\left[ |{H}_1|^2 \stackrel{x}{\circledast} |{H}_1|^2 \right] (x)= \left[ |{H}^{(l)}\out|^2 \stackrel{x}{\circledast}  |{H}^{(l)}\out|^2 \right ] (x-2x_0)
\end{equation}
If a Gaussian covariance model is assumed for aberrations, the auto-convolution $|{H}^{(l)}\out|^2 \stackrel{x}{\circledast}  |{H}^{(l)}\out|^2 $ should be maximum at $x=0$. In that case, the maximum position of $|{H}_1|^2 \stackrel{x}{\circledast} |{H}_1|^2(x) $ leads to an estimation of $2x_0$. Once the latter parameter is known, the undesired linear ramp can be removed from $\hat{\mathbf{U}}_1$ \eqref{phase_ramp}, as illustrated in Fig.~\ref{fig:correlationMatrix}(d). }

\rev{In the present case, this linear phase ramp compensation is applied to each estimated aberration phase law. Nevertheless, it should be noted that 
\lc{some specific shapes of aberrating layer -- such as a wedge -- can manifest} as a linear phase ramp in the aberration phase law. For such a  particular case, the lateral shift observed on the corresponding PSF is physical and should not be compensated \lc{for}. The removal of the linear phase ramp should thus be used with caution as it could cancel, in some specific cases, the benefits of the aberration correction process.}

\subsection{{Transmission matrix estimator}}

\subsubsection{Aberration correction in the receive transducer basis}
\al{Now that an estimator $\hat{\mathbf{U}}_1$ of \al{the aberration matrix} \rev{$\tilde{\mathbf{H}}^{(l)}\out$} has been derived, its phase conjugate can be used as a focusing law to compensate for aberrations. 
\la{We start by correcting in the transducer basis in receive (output). This means} 
building {an estimator $\mathbf{Q_{1}}$} of the transmission matrix from the Hadamard product 
\la{of} the free space transmission matrix $\mathbf{Q_{0}}$ and the phase conjugate of \al{$\hat{\mathbf{U}}_1$}:
\begin{equation}
    \al{\mathbf{Q_1} = \mathbf{Q_0} \circ \hat{\mathbf{U}}^*_1}.
    \label{eq:Q1_forcorrection}
\end{equation}
\la{$\mathbf{Q_1}$ is then used to recalculate the broadband \al{FR} matrix $\overline{\mathbf{R}}_{xx}{(z)}$ \eqref{eq:R_rr__time}, \wi{leading to} a corrected FR matrix $\mathbf{R}^{(c)}_{xx}(z)$ \rev{(Fig.~\ref{fig:schema__process_abCorr1}(b))}:} 
\begin{equation}
    \mathbf{R}^{(c)}_{xx}(z) = \mathbf{Q}_{1}^{\dagger}(z) \times \mathbf{Q_0}(z) \times \overline{\mathbf{R}}_{xx}(z).
    \label{eq:Rcorr_rr}
\end{equation}
Note that the correction is applied 
to the raw FR matrix ${\overline{\mathbf{R}}_{xx}(z)}$, and not to the filtered FR matrix $\al{\mathbf{R}'_{xx}(z)}$. 
\la{This is to make sure that} no 
\la{singly-scattered} echo is removed during the aberration correction process\lc{; as highly-aberrated singly-scattered echoes can extend quite far from the diagonal of the FR matrix, some of this information might be lost after a confocal filter is applied}.} 


\subsubsection{Aberration correction in the transmit plane wave basis}
\la{Next, 
aberration correction is performed for the transmit mode (input). 
It is important to note that it is possible to change the correction basis if need be. 
Here, 
correcting in the plane-wave basis in \wi{the transmit mode} is \rev{more efficient} since} 
the ultrasound emission sequence 
\la{was} performed in this basis. \al{Indeed, 
\la{
\al{it can} also help to} compensate for \rev{axial} movements of the medium that may have occurred during the acquisition. \rev{Such movements give rise to additional phase shifts in the illumination basis.} } \la{\al{\wi{A set of dual reflection matrices $\mathbf{R}'_{xk}(z)$}~\eqref{Rxk} is built from the updated FR matrix and an aberration phase law is extracted from the corresponding distortion matrices $\mathbf{D}'_{\wi{\mathbf{r}}k}(\rp)$~\cite{lambert2020distortion}.} The aberration correction process in the plane wave basis is similar to that described here for the transducer basis \eqref{eq:Q1_forcorrection}-\eqref{eq:Rcorr_rr}, replacing the matrix $\mathbf{Q}_0$ by $\mathbf{T}_0$~\eqref{T0} \rev{(Fig.~\ref{fig:schema__process_abCorr1}(c))}.} 
\la{
The result is} 
\al{an updated FR matrix $\mathbf{R}^{(c)}_{xx}(z)$\la{, an example of which} 
is displayed in \al{Fig.~\ref{fig:F}\rev{(e)}} at \rev{$z=39$~mm}. The corresponding CMP intensity profile is displayed in \al{Fig.~\ref{fig:F}\rev{(g)}} for the area $\mathcal{B}$ shown in \al{Fig.~\ref{fig:Full_results}\rev{(c)}}}. \al{Compared to the initial CMP intensity profile, the result of this first step of the aberration correction process seems quite modest (see \al{Table.}~\ref{tab:resolution-3dB}). This is explained by the relatively large size of the spatial window $\Delta \mathbf{r}$ chosen at the first step of the \al{UMI} process (Table.~\ref{tab:parameters}). While the central part of the FR matrix seems thinner in Fig.~\ref{fig:F}\rev{(e)} compared to its initial counterpart in Fig.~\ref{fig:F}\rev{(c)}, the focusing quality is not drastically improved on the eccentric parts of the field-of-view. \rev{These residual aberrations will be tackled in the next iteration of the aberration correction process by: (\textit{i}) projecting the ultrasound data in the transducer basis at input in order to compensate for aberrations induced by superficial layers at large depths; (\textit{ii}) projecting the ultrasound data in the plane wave basis at output in order to compensate for aberrations at shallow depths; (\textit{ii}) reducing the area $\Delta \mathbf{r}$ in order to address higher-order aberrations associated with smaller isoplanatic patches.}}
\begin{table}
    \centering
            \caption{\label{tab:resolution-3dB}\rev{Result of the {UMI} process in the area \rev{$\mathcal{B}$} {[Fig. \ref{fig:Full_results}(a)]}}.}
    \begin{tabular}{|c|c|c|c|c|c|}
        \hline
        Correction step & 0 & 1 & 2 & 3 & 4 \\
         \hline
       $F$ & 0.31 & 0.36 & 0.42 & 0.46 & 0.48 \\
        \hline
       $w_{-3dB}$ (mm) & 1.69 & 1.65 & 0.59 & 0.51 & 0.50 \\
         \hline
       Contrast  (dB) & -2.7 & -2.4 & -0.1 & 1.7 & 2.55 \\
        \hline
    \end{tabular}
\end{table}

\subsubsection{Iteration of the UMI process}

\al{The aberration correction process can now be iterated over smaller areas (see \al{Table.}~\ref{tab:parameters}). Note that the correction bases are also exchanged between input and output to minimize any redundancy in the algorithm and optimize the efficiency of this second step. 
Compared to the previous step, 
the CMP intensity profile displayed in Fig.~\ref{fig:F}(g) illustrates the gain both in terms of resolution and contrast of the input and output PSFs. \wi{After this second step,} the transverse resolution is actually enhanced by a factor \la{of} \rev{three} in the area $\mathcal{B}$ and the contrast shows an improvement of \rev{almost $3$~dB} [see \al{Table.}~\ref{tab:resolution-3dB}].\rev{ The transverse resolution is estimated from the full width at half maximum, $w_{-3dB}$, of the CMP intensity profile. The contrast 
is \wi{here computed} as the ratio between the single scattering energy and the incoherent background at focus. }}

\subsubsection{Convergence of the UMI process}

\al{The iteration of the aberration correction process can then be pursued since the quality of focus is improved at each step. 
\la{This} results in \la{more highly-resolved} 
virtual transducers 
\la{which}, in return, provides a better estimation of the aberration phase law (in particular at large angles in the plane wave basis or 
\la{near} the edge of the array in the transducer basis). 
As before, this process can be repeated at input and output in both correction bases, while \la{again} reducing 
the size of \la{the} isoplanatic patches \rev{and opening the confocal numerical pinhole} \la{(}
\al{Table.}~\ref{tab:parameters}). The \la{final} corrected FR matrix $\mathbf{R}^{(c)}_{xx}(z)$ 
is displayed in \al{Fig.~\ref{fig:F}\rev{(e)}}. 
\la{Comparing this result} with the initial $\overline{\mathbf{R}}_{xx}(z)$ [Fig.~\ref{fig:F}(b)] illustrates the benefit of \al{UMI}. \wi{Whereas the single scattering contribution originally spread over multiple resolution cells,} it now lies along the diagonal of the FR matrix\la{.} 
The part of the back-scattered energy that remains off-diagonal is mainly due to 
multiple scattering events taking place ahead of the focal plane~\cite{lambert2020reflection,lambert_ieee}. The corresponding CMP intensity profile is shown in Fig.~\ref{fig:F}\rev{(f)}. 
\la{A} comparison with the initial profile illustrates both the gain in terms of contrast \rev{($>5$ dB)} and resolution \rev{($\times 3.5$)} provided by \al{UMI} (see Table.~\ref{tab:resolution-3dB}).} 

\begin{figure}
    \centering 
    \includegraphics[width=1\columnwidth]{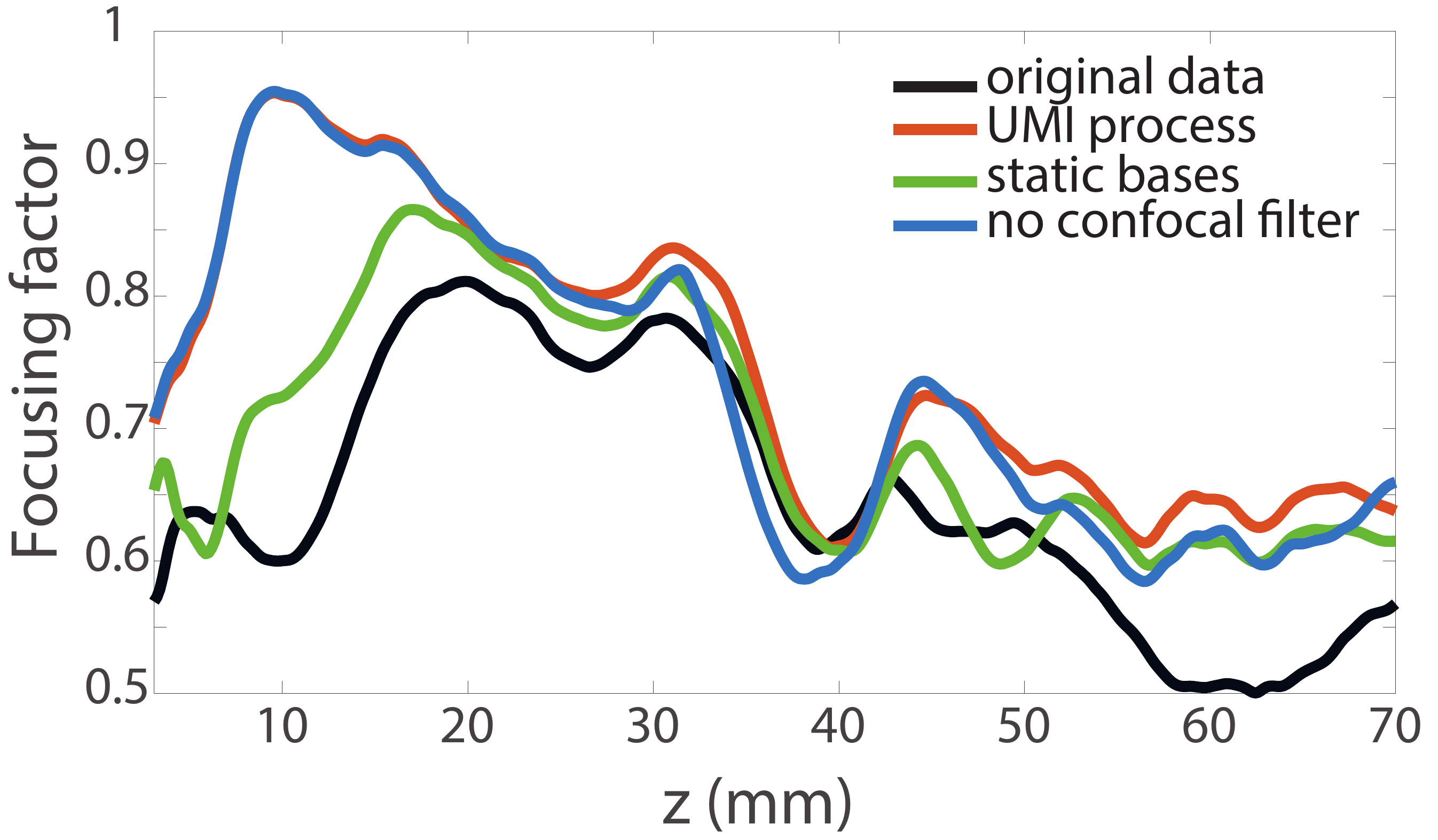}
    \caption{\rev{Depth evolution of the focusing factor $F$ for \revv{several aberration correction schemes applied to the gallbladder experiment}}.}
    \label{fig:F2}
\end{figure}
\rev{Figure~\ref{fig:F2} displays the depth evolution of the focusing factor $F$ obtained with the UMI process described above (red curve). It confirms the clear improvement of the focusing quality compared to its original value (black curve). It also shows the importance of alternating the correction bases (Fourier and transducer planes) both at input and output. If the aberration correction is only done in the measurement bases (plane wave basis at input and transducer basis at output), the focusing quality is much lower especially at shallow and large depths.  Finally, Fig.~\ref{fig:F2} highlights the effect of the adaptive confocal filter that improves the focusing quality, especially at large depths where the incoherent background 
\lc{dominates}. }

\rev{Nevertheless, note that the final focusing criterion reached by UMI ($F\sim 0.65$) remains far from its ideal value ($F= 1$). 
\lc{In the next section, we will show through numerical simulations that} this limit of the UMI process can be accounted for by the 
\lc{continued presence} of high-order aberrations. 
\lc{These aberration effects} are associated with an isoplanatic length smaller than the final size $\Delta \mathbf{r}$ of the spatial windows ({Table.}~\ref{tab:parameters}). Reducing 
$\Delta \mathbf{r}$ \lc{even further is not possible,} since the condition \eqref{con} would be no longer fulfilled, thereby leading to a biased estimation of the aberration phase laws. }

\section{Results} 

\subsection{{Numerical validation}}\label{par:num}
\begin{figure}
    \centering 
    \includegraphics[width=1\columnwidth]{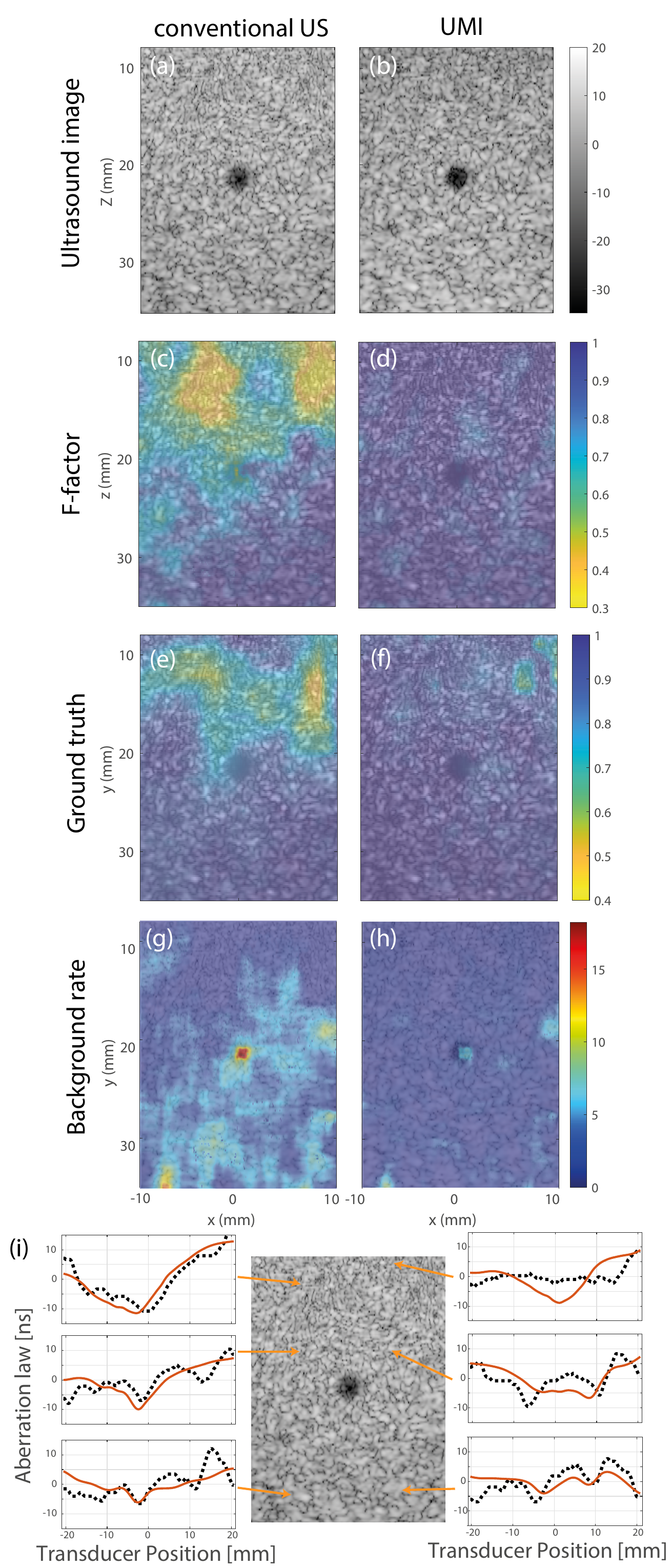}
    \caption{\rev{Results of the aberration correction process applied to the numerical experiment . (a) Conventional \rev{dynamically\lc{-}focused} image. (b) Corrected UMI image. Estimated (c,d) and true (e,f) $F-$maps before and after matrix correction of aberrations, respectively.} \rev{(g,h) Incoherent background rates before and after matrix correction of aberrations, respectively. (i) Examples of \revv{aberration laws} (red lines) in the transducer basis resulting from the matrix imaging process and compared to their true value (dashed black line).}}
    \label{fig:num1}
\end{figure}

\rev{
\lc{To validate our aberration correction method, and to investigate its limits, we now apply it to the} numerical simulation described in Ref.~\cite{lambert_ieee} (low speckle regime). 
\lc{For this simulation, the ultrasound transmission sequence used $161$ steering angles spanning from $-40^{o}$ to $40^{o}$.} The corresponding ultrasound image is displayed in Fig.~\ref{fig:num1}(a). 
\lc{The black, round inclusion near $z=20$~mm displays unexpected speckle noise -- this noise is caused by} aberrations induced by the inhomogeneous speed-of-sound and density distributions in superficial layers of tissues~\cite{lambert_ieee}. The F-map displayed in Fig.~\ref{fig:num1}(c) quantifies the aberration level in the ultrasound image. 
\lc{Un}surprisingly, the quality of focus is more degraded at short depths while aberrations are smoothed out at larger depths due to a smaller numerical aperture. This is confirmed by Fig.~\ref{fig:num1}(g) that shows several aberration laws across the field-of-view.  The previously described aberration correction process is applied using the parameters given in Tab.~\ref{tab:parameters}. The resulting ultrasound matrix image is displayed in Fig.~\ref{fig:num1}(b). Compared to the original image [Fig.~\ref{fig:num1}(a)], 
\lc{the corrected image} shows a clear contrast improvement in the anechoic inclusion. It can be quantified by the incoherent background rate \lc{[Fig.~\ref{fig:num1}(g,h)]} that is drastically decreased by the UMI process. A contrast improvement of 6 dB is observed in the vicinity of the anechoic inclusion. 
\lc{The transverse resolution is also vastly improved;} the $F-$factor now shows a nearly optimal value ($F\sim 1$) throughout the field-of-view 
[\lc{compare Figs.~\ref{fig:num1}(c) and (d)]. This} gain in focusing quality is confirmed by the ground-truth value of the $F-$factor 
\lc{[Fig.~\ref{fig:num1}(f)]} that is directly extracted from the local input and output PSFs, $H_\text{in}$ and $H_\text{out}$~\cite{lambert_ieee}. A slight discrepancy can be observed on top right part of the image and \rev{can be explained by the short-scale fluctuations of aberrations at shallow depths (see Fig.~\ref{fig:num1}(e)). As the focusing factor $F$ 
\lc{is} not able to grasp the short-scale variations of focusing quality in this area~\cite{lambert_ieee}, the spatial resolution of the transmission matrix estimator is probably not sufficient to capture the local aberration phase laws in this area.}}

\rev{The limits of UMI can also be highlighted by comparing the \revv{aberration} laws estimated by the UMI process with the true time delays. In the transducer basis, the \lc{latter} 
can actually be measured numerically by placing a source at a point of interest in the medium and recording the transmitted wave-front on the array of transducers. The comparison between the estimated and true time delay laws is shown for different locations in Fig.~\ref{fig:num1}(i). While UMI succeeds in retrieving the low spatial frequency components of the wave-front, its short-scale fluctuations are not captured by the aberration correction process. The main reason \lc{for this} is \lc{that} the 
\lc{local isoplanatic condition} 
required by the aberration correction procedure 
\lc{is} not satisfied by the high-order aberrations. \lc{Indeed, in} 
the transducer basis, the isoplanatic length is 
directly related the coherence length of the aberrating layer. The spatial resolution of the estimated aberration phase law is thus roughly given by the transverse size $\Delta_x$ (here 3 mm, see Tab.~\ref{tab:parameters}) of the spatial window considered for each local distortion matrix. The \revv{aberration} law is 
\lc{then} a low-pass filtered version of the true time delay law. This explains why $F$ does not reach a uniform and optimal value of 1 over the whole field-of-view after the aberration correction process 
\lc{[Fig.~\ref{fig:num1}(e)]}. Note also that the 
\lc{poor} estimation of the focusing law on the top right part of the image in Fig.~\ref{fig:num1}(i) is in agreement with the 
\lc{poor} focusing quality shown by Fig.~\ref{fig:num1}(f) in the same area.}

\begin{figure}
    \centering
    \includegraphics[width=1\columnwidth]{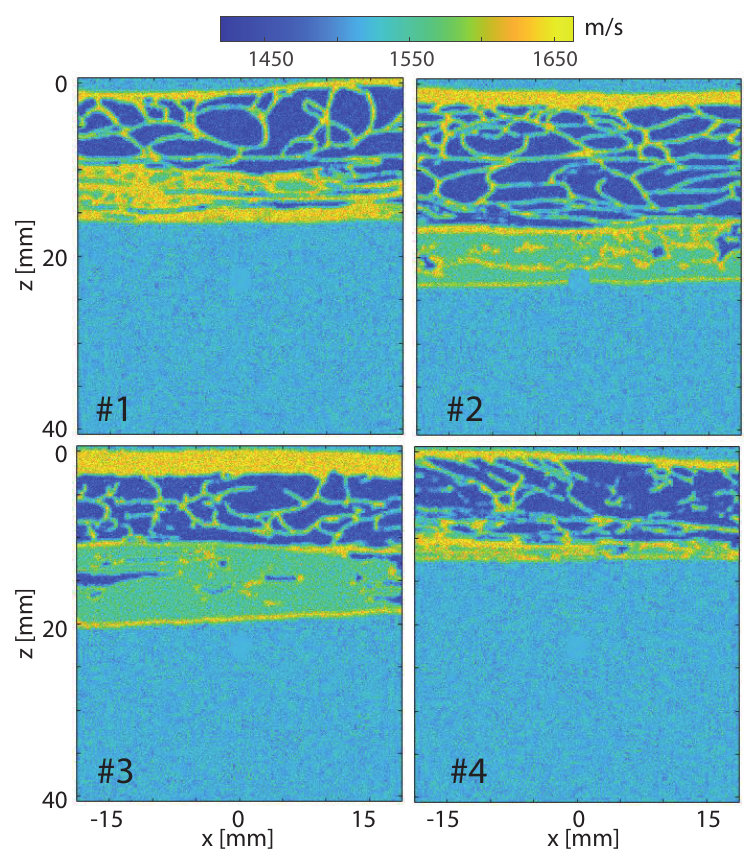}
    \caption{\revv{Speed-of-sound distributions for the different numerical models of the abdominal wall~\cite{Mast1997} simulated with k-wave.}}
    \label{fig:sos}
\end{figure}
\revv{To provide a more systematic study of UMI performance, we now consider a set of four numerical layers introduced by Mast \textit{et al.} in their seminal paper~\cite{Mast1997}. The corresponding speed-of-sound distributions are provided in Fig.~\ref{fig:sos}. For each realization, the level of aberration can be quantified by the Strehl ratio, $S$~\cite{mahajan1982strehl}. Initially introduced in the context of optical imaging, $S$ is defined as
the ratio of the peak intensity of the PSF with aberration to that without. Equivalently, it can also
be defined as the squared magnitude of the mean aberration phase law $\phi(\mathbf{u},\mathbf{r})$ in the transducer basis for a given point $\mathbf{r}$ in the medium: $S(\mathbf{r})=|\langle \exp \left \lbrace  i \phi (\mathbf{u},\mathbf{r}) \right \rbrace \rangle_{\mathbf{u}} |^2 $, where $\langle \cdots \rangle_{\mathbf{u}}$ denotes an average over the set of transducers $\mathbf{u}$. The Strehl ratio ranges from zero for a completely degraded focal spot to one for a perfect focusing. For each realization of the aberrating layer, the Strehl ratio $S$ has been measured and averaged over the speckle area behind the aberrating layer (20 mm$<z<$40 mm). This averaged value $\langle S (\mathbf{r})\rangle_{\mathbf{r}}$ is reported in Tab.~\ref{tab3}. Each realization shows a different Strehl ratio (hence aberration level) that depends on the arrangement of tissues in the simulated abdominal wall, the first realization being the most aberrating ($ \langle S \rangle \sim 0.5$).}

\revv{The UMI process is then applied using the parameters given in Tab.~\ref{tab:parameters}. The final Strehl ratio $S_{F}$ can be assessed from our estimator $\hat{\phi}(\mathbf{u},\mathbf{r}) $ of the aberration phase law:  $$S_{F}(\mathbf{r})=\left | \left \langle \exp \left \lbrace i [ \phi (\mathbf{u},\mathbf{r})-\hat{\phi} (\mathbf{u},\mathbf{r})]  \right \rbrace \right  \rangle_{\mathbf{u}} \right |^2 .$$}
\revv{Its averaged value $\langle S_F(\mathbf{r})\rangle_{\mathbf{r}}$ is reported in Tab.~\ref{tab3}. For each realization, a clear increase of the Strehl ratio is observed compared to its initial value $S$, the gain being higher when $S$ is smaller. For instance, this gain is drastic for realization~\#1: $\langle S_F \rangle / \langle S \rangle \sim $ 130 \%. Note also that the final Strehl ratio $S_F$ never exceeds a value of 0.9. As already highlighted by Fig.~\ref{fig:num1}(i), this is because UMI fails in capturing the short-scale fluctuations of the aberration phase laws.}
\begin{table}
    \centering
            \caption{\label{tab3}\revv{Result of the {UMI} process on four realizations of aberrating layers extracted from Ref.~\cite{Mast1997}}.}
    \begin{tabular}{|c|c|c|c|c|}
        \hline
        Realization & \#1 & \#2 & \#3 & \#4 \\
         \hline
       $\langle S \rangle$ & 0.49 & 0.8 & 0.79 & 0.76  \\
        \hline
       $\langle S_{F} \rangle$  & 0.64 & 0.87 & 0.89 & 0.83  \\
         \hline
    \end{tabular}
\end{table}

\subsection{Ultrasound matrix image}\label{par:results}

\lc{The UMI process now having been validated \revv{numerically}, we can turn our attention to the experimental ultrasound image $ \mathcal{I}_M$ which is built from the diagonal of the corrected FR matrices $R_{xx}^{(c)}$: $    \mathcal{I}_M(\mathbf{r})=  |R^{(c)}(\mathbf{r},\mathbf{r})|^2$.} 
\al{Fig.}~\ref{fig:Full_results} compares the resulting image $\mathcal{I}_M$ of the 
\lc{gallbladder} [Fig.~\ref{fig:Full_results}\al{(d)}] with the original one [Fig.~\ref{fig:Full_results}\al{(c)}]. The two images are normalized by their mean intensity and are displayed over the same dynamic range. A significant improvement of the image quality is observed especially at shallow depth and on the left of the gallbladder. 
\la{Figs.~\ref{fig:Full_results}\al{(c,d)} display\lc{s} \rev{the gallbladder in close-up}; UMI} reveals \lc{a clear view of the internal wall} 
that was completely blurred 
\lc{in} the original image. 
\lc{This improvement is a valuable one for in vivo imaging, as gallbladder inflammation commonly manifests as a thickening of the gallbladder wall.}


\lc{To quantify this improvement, maps of the focusing criterion are shown in Figs.~\ref{fig:Full_results}\al{(e,f)}, before and after the \al{UMI} process~\cite{lambert_ieee}}. The focusing parameter $F$ \rev{is improved} over a large part of the image [Fig.~\ref{fig:Full_results}\al{(f)}]. \rev{The gain in focusing quality is particularly spectacular at shallow depth\lc{s, as is evidenced by the increase in $F$ from $F\sim0.2$ [see Fig.~\ref{fig:Full_results}(e)] to an optimal value [$F\sim1$, see Fig.~\ref{fig:Full_results}(e)].} 
At larger depths, the F-factor is 
improved on \lc{both} the left and right sides of the gallbladder but does not reach an optimal value, saturating around $F\sim 0.7$. The subsistence of high-order aberrations \rev{associated with small isoplanatic lengths can account, at least partially, for this non-ideal focusing quality, even after the UMI process. An inhomogeneous attenuation of the wave-field across its angular spectrum can also explain this saturation of the $F-$factor, as 
\lc{will} be discussed below. } } 

\rev{To estimate the benefit of the UMI process in terms of image contrast, the background rate $\alpha$ 
\lc{was} estimated before and after the \al{UMI} process~\cite{lambert_ieee}. The corresponding $\alpha-$maps are shown in Figs.~\ref{fig:Full_results}\al{(g,h)}. The background rate $\alpha$ \rev{is drastically decreased} over a large part of the image [Fig.~\ref{fig:Full_results}\al{(h)}].  This is  particularly spectacular around the gallbladder which explains the much better quality of its close-up image (Fig.~\ref{fig:Full_results}(b))  compared to the original one (Fig.~\ref{fig:Full_results}(b)). 
\lc{The $\alpha-$map also shows that the contrast is less improved in a triangular region to the left of the image, and at larger depths ($z>40$ mm) behind the gall bladder. This effect} could be explained by an 
\lc{spatially-varying} 
attenuation across the angular spectrum of the incident wave-field\lc{,} induced} \rev{by upstream echogeneous structures. It would also account for the 
\lc{saturation} of the $F-$factor at large depths. 
\lc{Overcoming this issue would require an extension to the UMI process} in order to estimate both the amplitude and phase of the transmission matrix. This estimator would enable an inversion procedure of both the attenuation and aberrations induced by the medium rather than just a phase conjugation of wave-front distortions. }

\subsection{Estimated aberration phase laws} 
\label{phaselaw}

\begin{figure}[!ht]
    \centering 
    \includegraphics[width=1\columnwidth]{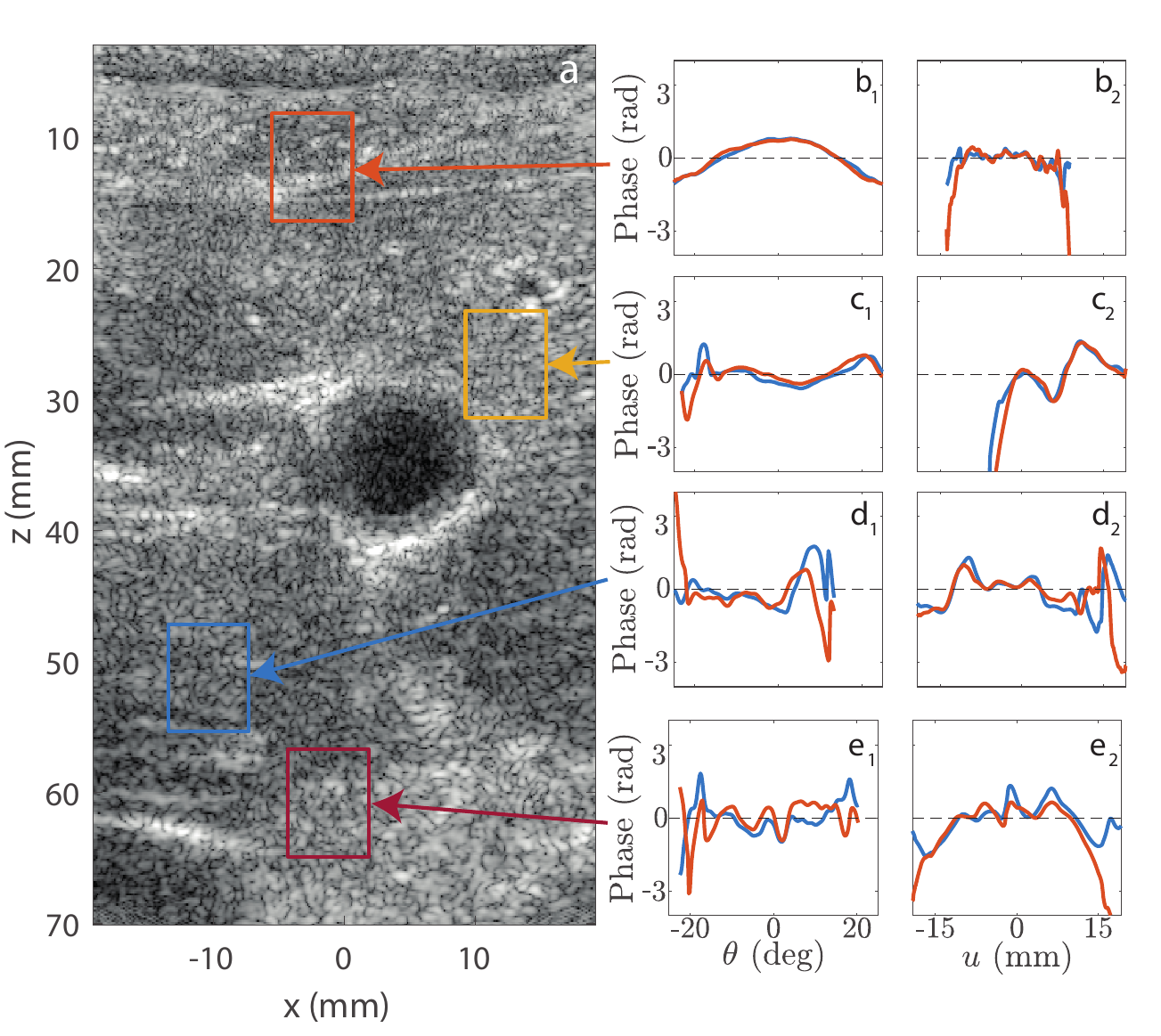}
    \caption{Examples of \rev{estimated} aberration \al{phase} laws resulting from the \al{matrix imaging} process, computed in transmit (blue curves) and in receive (red curves) modes, in the \al{plane wave (1) and transducer (2) bases}. The dimension of the selected areas is defined by the size of the spatial window used \al{at the fourth step of the whole process (see Table.1).}}
    \label{fig:phaseFullcorrection}
\end{figure}

\rev{Fig.~\ref{fig:phaseFullcorrection} shows the spatial distribution of \rev{estimated} \lc{aberration }
phase laws at the 
\lc{conclusion} of the matrix imaging process.
The truncated aspect of some of the aberration laws results from the maximal angles of illumination or collection imposed by the finite size of the ultrasonic array. At small depths [$z<A \tan [\beta (\mathbf{r})] /2$, see Fig.~\ref{fig:matrix_D}(k)], the spatial 
\lc{extent} of the reflected wave-field is limited by the numerical aperture of the probe [see e.g Fig.~\ref{fig:phaseFullcorrection}(b$_2$)]. A plane wave basis is thus 
\lc{preferable} since its angular range is almost invariant over the focal plane [see Fig.~\ref{fig:phaseFullcorrection}(b$_1$)]. On the contrary, in the far-field, the angular range of plane waves reaching each focal point is limited by the physical aperture of the probe [Fig.~\ref{fig:matrix_D}(k)]. 
\lc{The} transducer basis should \lc{thus} be favoured 
\lc{at large depths} [see e.g \ref{fig:phaseFullcorrection}(d)].}

\rev{Fig.~\ref{fig:phaseFullcorrection} highlights the spatial variations of the aberration phase across the field-of-view. 
\lc{This} spatial resolution 
\lc{was} made possible by gradually reducing the size of the spatial window $W_{\Delta \mathbf{r}}$ at each step of the matrix imaging process (see Table.~\ref{tab:parameters}). The four colored straight rectangles in Fig.~\ref{fig:Full_results}(c) \al{depict} the size of the spatial windows used at each step. \al{A $75\%$ \lc{overlap }
is applied between 
\lc{subsequent spatial windows} in order to retrieve the aberration matrix $\tilde{\mathbf{H}}_\textrm{in/out}$ at a high resolution $\Delta \mathbf{r}$.}  The spatial correlations in the aberration matrices 
\lc{can now be quantitatively investigated (Sec.\ \ref{par:aberrationLaw}).}} 

\subsection{{Aberration matrices and isoplanatic modes}} 

\begin{figure}
    \centering 
    \includegraphics[width=1\columnwidth]{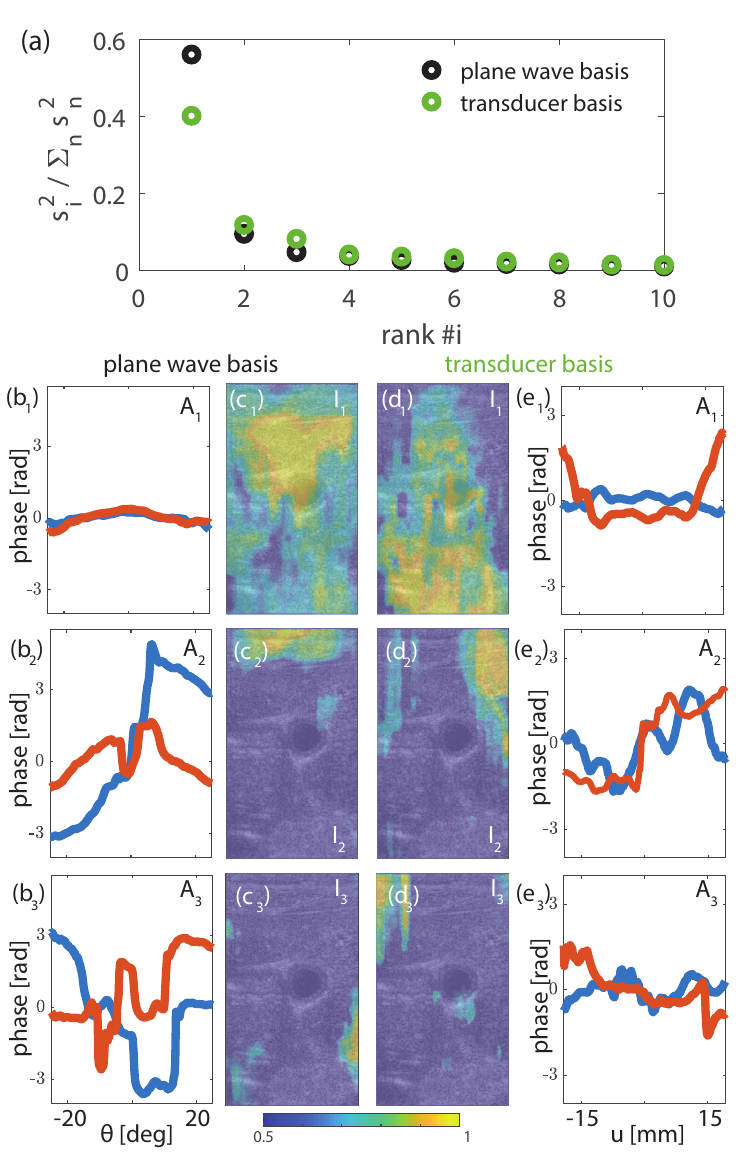}
    \caption{\rev{(a)  First eigenvalues of the correlation matrix $\mathbf{C}^{(H)}$ in the plane wave and transducer bases. (b)} Phase of the \al{first three} singular vectors $\mathbf{A}_p$ (from top to bottom) in the plane wave basis (blue line: $\mathbf{A}\in$, red line: $\mathbf{A}\out$) and \wi{(b)} corresponding isoplanatic \rev{modes} $\mathbf{I}_p$. \rev{(d)} Phase of the first \al{three} singular vectors $\mathbf{A}_p$ (from top to bottom) in the transducer basis (blue line: $\mathbf{A}\in$, red line: $\mathbf{A}\out$)\lc{. } 
    \rev{(c)} corresponding isoplanatic \rev{modes} $\mathbf{I}_p$. 
    }
    \label{fig:IP}
\end{figure}


Matrix imaging also enables 
\la{the} mapping of \rev{isoplanatic modes} across the field-of-view. 
\la{\rev{As already shown for the distortion matrix with specular reflectors~\cite{Badon2019},} a SVD \rev{can be} highly useful in extracting the characteristic spatial variations of aberration.} 
\rev{To that aim, we define a full aberration matrix from the input and output aberration matrices defined either in the plane wave or} \rev{transducer bases:
\begin{equation}
   \tilde{\mathbf{H}}=\left [\begin{array}{c} \tilde{\mathbf{H}}\in \\ \tilde{\mathbf{H}}\out  \end{array} \right ]. 
\end{equation}
The input and output aberration phase laws, estimated for each point $\rp$, are stored along each column of $\tilde{\mathbf{H}}=[\tilde{{H}}(\lbrace v\in,v\out \rbrace,\rp)]$, with $v=u$ or $k$ depending on the considered basis.} \rev{The SVD of $\tilde{\mathbf{H}}$ can be written as follows:}
\begin{equation}
   \tilde{\mathbf{H}} =  \sum_p s_p \mathbf{A}_{p}  \times \mathbf{I}_p^\dag .
\end{equation}
\rev{where $s_p$ are the singular values \lc{ar}ranged in decreasing order. $\mathbf{A}_{p}=[{A}_{p}(\lbrace v\in,v\out \rbrace)]$ are the singular vectors in the transducer or Fourier bases
\lc{, and} $\mathbf{I}_{p}=[\mathbf{I}_{p}(\rp)]$ are the singular vectors in the focused basis.} 
\rev{
\lc{For a physical interpretation of these vectors, we take advantage of the equivalence between the SVD of $\tilde{\mathbf{H}}$ and }
the eigenvalue decomposition of the spatial correlation matrix, $\mathbf{C}^{(H)}=\tilde{\mathbf{H}}^{\dag} \times \tilde{\mathbf{H}}$
. The elements of 
\lc{$\mathbf{C}^{(H)}$} correspond to the correlation coefficients between aberration phase laws obtained for each image pixel $\rp$ and $\mathbf{r}'_p$: $$C^{(H)}(\mathbf{r}'_p,\rp)= \sum_{\lbrace v\in,v\out \rbrace}\tilde{{H}}^*(\lbrace v\in,v\out \rbrace,\mathbf{r}'_p){\tilde{{H}}(\lbrace v\in,v\out \rbrace,\rp)} .$$ The first eigenvector $\mathbf{I}_1$ is \lc{thus} the spatial domain where the degree of correlation between aberration phase laws is maximized. This degree of correlation is quantified by the normalized eigenvalue $\hat{s}_1^2$, such that 
\begin{equation}
\hat{s}_p^2=\frac{s_p^2}{\sum_i s_i^2}=\frac{\mathbf{I}_p^{\dag} \times \mathbf{C}^{(H)} \times \mathbf{I}_p}{\mbox{Tr} [ \mathbf{C}^{(H)}]}.
\end{equation}
\rev{The corresponding singular vector $\mathbf{A}_1=s_1^{-1}\tilde{\mathbf{H}} \times \mathbf{I}_1$ contains the associated input and output aberration phase laws, \rev{$\mathbf{A}_{\textrm{in},1}=[{A}_{\textrm{in},1}( v\in)]$ and $\mathbf{A}_{\textrm{out},1}=[{A}_{\textrm{out},1}( v\out)] $}, such that  \begin{equation*}
    \mathbf{A}_{p}=\left [\begin{array}{c} \mathbf{A}_{\textrm{in},p} \\ \mathbf{A}_{\textrm{out},p}  \end{array} \right ] .
\end{equation*} 
The same process can be iterated on the matrix {$\tilde{\mathbf{H}}-s_1  \mathbf{A}_{1}  \times \mathbf{I}_1^\dag$} to retrieve the second eigenstate and so on. A set of orthogonal isoplanatic \rev{modes} $\mathbf{I}_p$ is finally obtained with a degree of correlation $\hat{s}_p$ that decreases with their rank.}}

\rev{Fig.~\ref{fig:IP}(a) displays the corresponding eigenvalues in the plane wave and transducer bases. A few predominant eigenvalues associated with the main isoplanatic modes seem to emerge from a continuum of lower eigenvalues associated with a noise background in each case. To determine the effective number of isoplanatic \rev{modes} supported by the field-of-view in each basis, one can consider the entropy \rev{$\mathcal{H}(\hat{s}_p^2)$ of the normalized eigenvalues $\hat{s}_p^2 $ that yields the effective rank of the matrix $\mathbf{C}^{(H)}$:
$   \mathcal{H}(\hat{s}_p) = -\sum_i \hat{s}_i \log_2 \left( \hat{s}_i \right )$. Here, the entropy is $3.1$ and $3.9$ in the Fourier and transducer bases, respectively.}}

\al{Fig.~\ref{fig:IP} shows the three first eigenstates of \rev{$\tilde{\mathbf{H}}$} in the plane wave and transducer bases. 
\la{We first remark} that the retrieved aberration phase laws, $\mathbf{A}_{\textrm{in},p}$ and $\mathbf{A}_{\textrm{out},p}$, are not strictly equal at input and output\la{; this is despite the fact that} 
the transmit and back-scattered waves travel through the same heterogeneities\la{, and so the} 
manifestation of aberrations should 
be \al{identical} 
\la{in} input and output.} 
\al{The partial non-reciprocity stems from the different input and output bases used to: (\textit{i}) originally record the reflection matrix; (\textit{ii}) correct aberrations at each step of the \al{UMI} process [see Table.~\ref{tab:parameters}]. For this latter reason, some distorted components can emerge \la{more clearly} 
in one basis 
\la{rather than} the other, as aberrations in each basis are not fully independent (especially in the far-field).
} 

\al{Interestingly, the \la{measured features of an isoplanatic \rev{mode}} 
differ according to the correction basis. In particular, the \rev{first isoplanatic mode $\mathbf{I}_1$ [Figs.~\ref{fig:IP}(c$_1$,d$_1$)]} confirms the fact that the plane wave and transducer bases are more effective at small and large depths, respectively \rev{[see Sec.~\ref{phaselaw}]}. Moreover, each correction basis addresses aberrations of \la{a} different nature. Indeed, the spatial extension of the isoplanatic patches is deeply affected by the correction basis, thereby impacting the result of the aberration correction process. 
Depending on the location of the aberrating layer and/or its spatial dimension, one basis 
\la{is therefore} more suitable than 
\la{another} to extract the aberration law. }

\rev{On the one hand,} in most \textit{in-vivo} applications, the organs under study are generally separated from the probe by 
\la{layers of skin, adipose and/or muscle tissues. In such 
layered media}, aberrations are invariant by translation from the plane wave basis. \rev{The plane wave eigenstates displayed in Fig.~\ref{fig:IP} confirm this statement. \rev{While $\mathbf{I}_2$ focuses on the superficial layers of skin \rev{and fat}  [Fig.~\ref{fig:IP}(c$_2$)], $\mathbf{I}_1$ mainly spreads between these superficial layers and the gallbladder [Fig.~\ref{fig:IP}(c$_1$)]. The corresponding aberration phase laws exhibit a parabolic shape with different curvatures, which is characteristic of a layered medium [Figs.~\ref{fig:IP}(b$_1$,b$_2$)].}}

\rev{On 
other hand,} a local perturbation of the medium speed-of-sound located at shallow depth such as superficial veins 
will 
\la{have a strong impact on the signals that are measured by transducers located directly} 
above. The transducer basis is then the most adequate for those local variations of the speed-\la{of-}sound. \rev{$\mathbf{I}_2$ and $\mathbf{I}_3$ focus on the top right and left parts of the ultrasound image, respectively 
[see Figs.~\ref{fig:IP}(d$_2$,d$_3$)]. The corresponding aberration phase laws exhibit fluctuations of high spatial frequencies\lc{; such phase laws are} characteristic \la{of those which would be induced by local speed-of-sound variations at shallow depths} 
[Fig.~\ref{fig:IP}(e$_2$,e$_3$)].}

\lc{Interestingly, the overall observation is a correlation between the calculated isoplanatic \rev{modes} and the tissue architecture \wi{revealed by UMI} [Fig.~\ref{fig:Full_results}\rev{(d)]}. 
The SVD of the aberration matrices thus provides a segmentation of ultrasound images that 
\rev{could be related to} the actual distribution of tissues in the field-of-view. This information 
\lc{would be valuable for} the quantitative mapping of 
\lc{important} bio-markers such as the speed-of-sound~\cite{Imbault2017,lambert2020reflection,Jaeger2021}.}


\section{Discussion and Perspectives}

\al{
\lc{In this paper, we have presented a validation of} the UMI process for local aberration correction \rev{by means of a numerical simulation of aberrations through the abdominal wall\lc{,} and an \textit{in vivo} experiment performed on a human gallbladder. Ultrasound imaging is actually often used for diagnosis of cholecystitis~\cite{OextquotesingleConnor2011} (inflammation of the gallbladder)\lc{, either by} 
detecting gallstones or identifying pericholic fluid and thickened gallbladder wall.} 
\lc{This application of \textit{in-vivo} ultrasound imaging is one of the most challenging, involving} 
\la{various types} of tissues and a strongly heterogeneous distribution of the speed-of-sound.} 
\la{
\lc{Such a medium} includes areas of both strong and weak scattering, which \wi{is a 
\lc{difficult} problem}} \la{for standard adaptive focusing techniques. 
\lc{B}ack-scattered echoes are generated either by unresolved scatterers (ultrasound speckle), bright point-like scatterers, anechoic regions (gallbladder) 
\la{or specular} \al{structures} that are 
\la{larger than the image resolution (for example, the skin-muscle 
interface around $z=5$~mm)}. Previous works have shown that the \al{distortion matrix concept can be applied to both specular objects~\cite{Badon2019}, random scattering media~\cite{lambert2020distortion}, or sparse media made of a few isolated scatterers~\cite{Touma2020}}. 
\lc{This} article details \al{
\la{a} global strategy} 
\la{for aberration correction} that is essential for applying \al{UMI} to \textit{in-vivo} \al{configurations} where \al{all \la{of} these scattering} regimes can be found.} 

While our method is inspired by previous works in ultrasound imaging\al{~\cite{ODonnell1988,Mallart1994,varslot2004eigenfunction,robert2008green,montaldo2011time}}, it features several distinct and important differences. The first one is its primary building block: The broadband \al{FR} matrix that precisely selects all \la{of} the 
\la{singly-scattered} echoes originating from each focal point. 
\la{This} is a decisive step since it 
\la{greatly} reduces the contribution of\rev{: (\textit{i}) out-of-focus echoes thanks to ballistic time gating; (\textit{i})  multiply-scattered echoes by means of an adaptive confocal filter.} 
Secondly, the matrix approach provides a generalization of the virtual transducer interpretation~\cite{robert2008green}. 
By decoupling the location of the input and output focal spot\la{s}, this approach becomes flexible enough for a \rev{local estimation and compensation of aberrations} at both input and output and in any correction basis, thereby unifying the different methods developed in the past for aberration correction from a single basis~\cite{robert2008green,montaldo2011time,Jaeger2015SosMap,chau2019locally,bendjador2020svd}.

\rev{On the one hand, one can easily project the FR matrix at input (or output) to build a $\mathbf{D}$-matrix in the different bases of interest by simply considering monochromatic propagators at the central frequency. On the other hand, the output (or input) of the $\mathbf{D}$-matrix remains in a focused basis which enables a local estimation of aberration phase laws by truncating the field-of-view into limited spatial windows. It thus enables an iterative procedure of aberration correction with alternative changes of correction basis between input and output, while gradually improving the spatial resolution of the transmission matrix estimator. Last but not least, local aberration phase laws are estimated in the Fourier domain, which is much more efficient in terms of computational speed~\cite{chau2019locally} than the temporal cross-correlation of time-delayed signals usually performed for adaptive focusing. It also enables the estimation of the aberration phase laws 
\lc{via} a SVD of the distortion matrix~\cite{lambert2020distortion,bendjador2020svd}. Equivalent to an iterative time reversal process~\cite{varslot2004eigenfunction,robert2008green,montaldo2011time}, this estimator is more robust than usual correlation techniques~\cite{flax1988phase,Jaeger2015SosMap,chau2019locally} with respect to \lc{guide-star blurring}.} 

\rev{UMI thus} constitutes a powerful tool for imaging a heterogeneous medium when little to no previous knowledge on the spatial variations of the speed-of-sound is available. \rev{It can be applied whatever the array geometry (curved probes, phased arrays, matrix arrays, \textit{etc.}) 
\lc{or} acquisition scheme (focused excitations, plane wave, diverging waves, \textit{etc.}).} 
\al{Optimized contrast and resolution can be recovered for any pixel of the ultrasound image. 
\la{This approach} also paves the way towards a mapping of the speed-of-sound distribution inside the medium~\cite{Imbault2017,lambert2020reflection,Jaeger2021} by revealing the different isoplanatic \rev{modes} in the ultrasound image}. \al{Such an optimized focusing process 
\la{is also critical for accurate characterization measurements such as local measurements of ultrasound} 
attenuation~\cite{Suzuki1992}, 
scattering anisotropy~\cite{Rodriguez-Molares2017} or the micro-architecture of soft tissues~\cite{Franceschini2012}.} 

\section*{Acknowledgment}
\rev{The authors wish to thank Paul Balondrade and Ulysse Najar whose own research works in optics inspired this study, as well as anonymous reviewers whose feedback has helped us to improve the quality of the manuscript.}

%

\end{document}